%% file: main.tex
\PassOptionsToPackage{prologue,usenames,dvipsnames,table}{xcolor}
\PassOptionsToPackage{bookmarks,unicode,colorlinks=true}{hyperref}

\documentclass[sigconf]{acmart}

\input{z-header}

\AtBeginDocument{%
  }

\makeatletter
\renewcommand\paragraph{\@startsection{paragraph}{4}{\z@}%
  {-.5\baselineskip \@plus -2\p@ \@minus -.2\p@}%
  {-3.5\p@}%
  {\ACM@NRadjust{\@parfont\@adddotafter}}}
\makeatother

\setcopyright{acmlicensed}
\copyrightyear{2024}
\acmYear{2024}
\acmDOI{10.1145/3691620.3695487}

\acmConference[ASE '24]{the 39th IEEE/ACM International Conference on Automated Software Engineering}{October 27--November 1, 2024}{Sacramento, California, United States}
\acmISBN{978-1-4503-XXXX-X/18/06}

\begin{document}

\title{\textsc{Parf}: Adaptive Parameter Refining for Abstract Interpretation}

\author{Zhongyi Wang}
\orcid{0009-0008-1986-6070}
\authornote{Both authors contributed equally to this research.}
\affiliation{%
  \institution{Zhejiang University}
  \city{Hangzhou}
  \country{China}}
\email{wzygomboc@zju.edu.cn}

\author{Linyu Yang}
\orcid{0009-0007-3838-0538}
\authornotemark[1]
\affiliation{%
  \institution{Zhejiang University}
  \city{Hangzhou}
  \country{China}}
\email{linyu.yang@zju.edu.cn}

\author{Mingshuai Chen}
\authornote{Corresponding author.}
\orcid{0000-0001-9663-7441}
\affiliation{%
  \institution{Zhejiang University}
  \city{Hangzhou}
  \country{China}}
\email{m.chen@zju.edu.cn}

\author{Yixuan Bu}
\orcid{0009-0000-4720-9758}
\affiliation{%
  \institution{Zhejiang University}
  \city{Hangzhou}
  \country{China}}
\email{yixuanbu@zju.edu.cn}

\author{Zhiyang Li}
\orcid{0009-0000-5632-9479}
\affiliation{%
  \institution{Zhejiang University}
  \city{Hangzhou}
  \country{China}}
\email{misakalzy@zju.edu.cn}

\author{Qiuye Wang}
\orcid{0000-0001-5138-3273}
\affiliation{%
  \institution{Fermat Labs, Huawei Inc.}
  \city{Dongguan}
  \country{China}}
\email{wangqiuye2@huawei.com}

\author{Shengchao Qin}
\orcid{0000-0003-3028-8191}
\affiliation{%
  \institution{Xidian University}
  \city{Xi'an}
  \country{China}}
\email{shengchao.qin@gmail.com}

\author{Xiao Yi}
\affiliation{%
  \institution{Fermat Labs, Huawei Inc.}
  \city{Hong Kong}
  \country{China}}
\email{yi.xiao1@huawei.com}

\author{Jianwei Yin}
\orcid{0000-0003-4703-7348}
\affiliation{%
  \institution{Zhejiang University}
  \city{Hangzhou}
  \country{China}}
\email{zjuyjw@zju.edu.cn}

\renewcommand{\shortauthors}{Wang et al.}

\input{body/abstract}



\maketitle

\setlength{\floatsep}{1\baselineskip}
\setlength{\textfloatsep}{1\baselineskip}
\setlength{\intextsep}{1\baselineskip}

\input{body/introduction}
\input{body/background}
\input{body/problem_statement}
\input{body/methodology}
\input{body/experiments}
\input{body/limitations}
\input{body/relatedwork}
\input{body/conclusion}

\begin{acks}
    %
    This work has been funded by the ZJNSF Major Program under grant No.~LD24F020013, by the CCF-Huawei Populus Grove Fund under grant No.~CCF-HuaweiFM202301, by the Fundamental Research Funds for the Central Universities of China under grant No.~226-2024-00140, and by the ZJU Education Foundation's Qizhen Talent program. The authors would like to thank Shenghua Feng for helpful discussions and the anonymous reviewers for their constructive feedback.
\end{acks}

\bibliographystyle{ACM-Reference-Format}
\bibliography{references}
\end{document}

%% file: z-header.tex
\usepackage{algorithm}
\usepackage{algorithmic}
\usepackage{todonotes,xspace}

\usepackage{graphicx}
\usepackage{caption}
\usepackage{subcaption}
\usepackage{amsmath}
\usepackage{mathtools}
\usepackage{csquotes}

\usepackage{fancyvrb}
\usepackage[frozencache]{minted}
\setminted[c]{
    frame=lines,
    framesep=2mm,
    baselinestretch=1.2,
    fontsize=\footnotesize, 
    breaklines=true,
    xleftmargin=0em,
    xrightmargin=0em
}
\usepackage{framed}

\usepackage{adjustbox}
\usepackage{makecell}
\usepackage{tikz}
\usetikzlibrary{patterns,calc,arrows,shapes,snakes,automata,backgrounds,petri,positioning,decorations.pathmorphing,intersections}
\usepackage{tkz-euclide}
\tikzset{mynode/.style={draw,solid,circle,inner sep=1pt}}
\usepackage[customcolors]{hf-tikz}
\hfsetfillcolor{gray!10}
\hfsetbordercolor{none}
\usepackage{pgfplots}
\usepgfplotslibrary{units}  
\pgfplotsset{compat=1.17}
\usepackage{multirow}

\usepackage[capitalise]{cleveref}

\newcommand{\green}[1]{\textcolor{ForestGreen}{#1}}
\newcommand{\blue}[1]{\textcolor{blue}{#1}}

\newcommand{\orange}[1]{\textcolor{RedOrange}{#1}}

\newcommand{\maroon}[1]{\textcolor{Maroon}{#1}}

\newcommand{\best}[1]{{\textbf{#1}}}
\newcommand{\similar}[1]{{\underline{#1}}}

\newcommand{\default}{%
\textsc{Default}\xspace}
\newcommand{\expert}{%
\textsc{Expert}\xspace}
\newcommand{\official}{%
\textsc{Official}\xspace}
\newcommand{\parfstrategy}{%
\parf\xspace}

\newcommand{\parf}{\textsc{Parf}\xspace}
\newcommand{\eva}{\textsc{Frama-C/Eva}\xspace}
\newcommand{\framac}{\textsc{Frama-C}\xspace}
\newcommand{\framacsv}{\textsc{Frama-C-SV}\xspace}
\newcommand{\symtuner}{\textsc{SymTuner}\xspace}
\newcommand{\mopsa}{\textsc{Mopsa}\xspace}

\newcommand{\randomvariable}[3]{#1^{#2}_{\textnormal{#3}}}
\newcommand{\randomvector}[2]{#1_{\textnormal{#2}}}

\newcommand{\procedure}[1]{\ensuremath{\mathtt{#1}}}

\newcommand{\Nats}{\ensuremath{\mathbb{N}}\xspace}

\newcommand{\PosIntsInf}{\mathbb{Z}_{\geq 0}^\infty\xspace}

\newcommand{\StrictPosReals}{\mathbb{R}_{> 0}}

\newcommand{\bTRUE}{\ensuremath{\textit{true}}}
\newcommand{\bFALSE}{\ensuremath{\textit{false}}}

\newcommand{\defeq}{{}\triangleq{}}
\newcommand{\ddefeq}{\ {}\defeq{}\ }
\newcommand{\eeq}{~{}={}~}
\newcommand{\ssim}{~{}\sim{}~}

\newcommand{\qoplus}{\quad{}\oplus{}\quad}

\newcommand{\tto}{~{}\to{}~}

\newcommand{\qeq}{\quad{}={}\quad}

\newcommand{\ssqsubseteq}{~{}\sqsubseteq{}~}
\newcommand{\ssubseteq}{~{}\subseteq{}~}

\newcommand{\qiimplies}{\quad\textnormal{implies}\quad}

\def\triangleforqed{\hbox{$\lhd$}}
\makeatletter
\DeclareRobustCommand{\qedT}{%
	\ifmmode
	\eqno \def\@badmath{$$}
	\let\eqno\relax \let\leqno\relax \let\veqno\relax
	\hbox{\triangleforqed}%
	\else
	\leavevmode\unskip\penalty9999 \hbox{}\nobreak\hfill
	\quad\hbox{\triangleforqed}%
	\fi
}
\makeatother

\newtheoremstyle{rmkstyle}{}{}{}{}{\bfseries}{.}{ }{\thmname{#1}\thmnote{ (#3)}}
\theoremstyle{rmkstyle}

\newtheorem{remark}{Remark}

\newcommand{\revision}[1]{#1}

%% file: body/abstract.tex
\begin{abstract}
Abstract interpretation is a key formal method for the static analysis of programs. The core challenge in applying abstract interpretation lies in the configuration of abstraction and analysis strategies encoded by a large number of external parameters of static analysis tools. To attain low false-positive rates (i.e., accuracy) while preserving analysis efficiency, tuning the parameters heavily relies on expert knowledge and is thus difficult to automate. In this paper, we present a fully automated framework called {\parf} to adaptively tune the external parameters of abstract interpretation-based static analyzers. {\parf} models various types of parameters as random variables subject to probability distributions over latticed parameter spaces. It incrementally refines the probability distributions based on accumulated intermediate results generated by repeatedly sampling and analyzing, thereby ultimately yielding a set of highly accurate parameter settings within a given time budget. We have implemented {\parf} on top of {\eva} -- an off-the-shelf open-source static analyzer for C programs -- and compared it against the expert refinement strategy and {\eva}'s official configurations over the {\framac} OSCS benchmark. Experimental results indicate that {\parf} achieves the lowest number of false positives on 34/37 (91.9\%) program repositories with exclusively best results on 12/37 (32.4\%) cases. In particular, {\parf} exhibits promising performance for analyzing complex, large-scale real-world programs.

\end{abstract}

%% file: body/introduction.tex
\section{Introduction}\label{sec:intro}

The theory of abstract interpretation -- ever since its inception in the 1970s by Cousot and Cousot~\cite{cousotAbstractInterpretationUnified1977} -- has witnessed significant applications in the field of static program analysis, which aims to identify potential runtime errors (RTEs) without executing the program. 
Due to its core mechanism of safely approximating the concrete program semantics, abstract interpretation-based static analysis features soundness, i.e., true alarms of RTEs will not be missed, yet not completeness, i.e., false alarms may be emitted. These false alarms do not induce RTEs and thus may be eliminated by conducting more accurate approximations at the cost of less efficient analysis. State-of-the-art static analysis tools, such as Astr{\'e}e~\cite{kastnerAbstractInterpretationIndustry2023}, {\framac}~\cite{DBLP:journals/fac/KirchnerKPSY15}, \textsc{Goblint}~\cite{saanGoblintThreadModularAbstract2021}, \mopsa~\cite{DBLP:conf/vstte/JournaultMMO19}, and \textsc{Sparrow}~\cite{10.1145/2254064.2254092}, integrate multiple abstraction and analysis strategies encoded by various external parameters, thereby enabling users to balance analysis accuracy and efficiency by tuning these parameters.

\begin{figure}[t]
  \centering
  \scriptsize
  \begin{Verbatim}[commandchars=\\\{\}]
[eva] Option \textbf{-eva-precision 3} detected, automatic configuration of the analysis:
  option -eva-min-loop-unroll set to \orange{0} (default value).
  option -eva-auto-loop-unroll set to \orange{64}.
  option -eva-widening-delay set to \orange{2}.
  option -eva-partition-history set to \orange{0} (default value).
  option -eva-slevel set to \orange{35}.
  option -eva-ilevel set to \orange{24}.
  option -eva-plevel set to \orange{70}.
  option -eva-subdivide-non-linear set to \orange{60}.
  option -eva-remove-redundant-alarms set to \blue{true} (default value).
  option -eva-domains set to \green{'cvalue,equality,gauges,symbolic-locations'}.
  option -eva-split-return set to \maroon{''} (default value).
  option -eva-equality-through-calls set to \maroon{'none'}.
  option -eva-octagon-through-calls set to \blue{false} (default value).
  \end{Verbatim}
  \caption{A typical parameter setting (under precision 3) of {\eva}~\cite{eva_user_manual} with different parameter types: \orange{integer}, \blue{Boolean}, \maroon{string}, 
  and \green{set-of-strings}.}
  \label{fig:parameter-setting-precision3}
\end{figure}

\begin{figure*}[t]
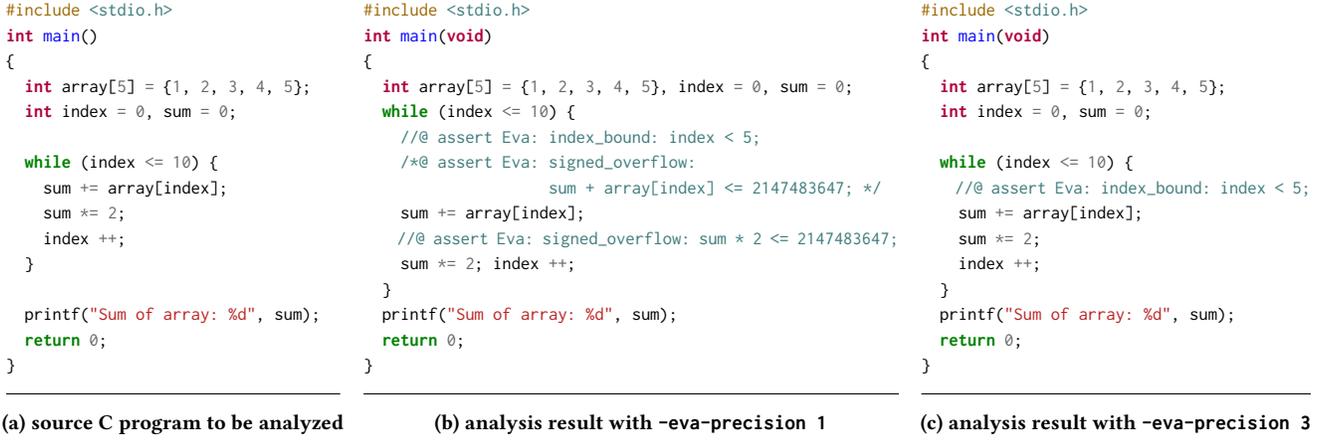

	\centering
		\begin{subfigure}[b]{0.26\linewidth}
			\centering
      \begin{minipage}{0.96\textwidth}
\begin{minted}{c}
#include <stdio.h>
int main()
{
  int array[5] = {1, 2, 3, 4, 5};
  int index = 0, sum = 0;
  
  while (index <= 10) {
    sum += array[index];
    sum *= 2;
    index ++;
  }
  
  printf("Sum of array: %d", sum);
  return 0;
}
\end{minted}
    \end{minipage}
			\caption{source C program to be analyzed}
			\label{fig:source-prog}
		\end{subfigure}
		\hfil
		\begin{subfigure}[b]{0.4\linewidth}
			\centering
   \begin{minipage}{1\textwidth}
\begin{minted}{c}
#include <stdio.h>
int main(void)
{
  int array[5] = {1, 2, 3, 4, 5}, index = 0, sum = 0;
  while (index <= 10) {
    //@ assert Eva: index_bound: index < 5;
    /*@ assert Eva: signed_overflow:
                    sum + array[index] <= 2147483647; */
    sum += array[index];
    //@ assert Eva: signed_overflow: sum * 2 <= 2147483647;
    sum *= 2; index ++;
  }
  printf("Sum of array: %d", sum);
  return 0;
}
    \end{minted}
    \end{minipage}
			\caption{analysis result with \texttt{-eva-precision 1}}
			\label{fig:low-precision}
		\end{subfigure}
		\hfil
		\begin{subfigure}[b]{0.3\linewidth}
			\centering
   \begin{minipage}{.97\textwidth}
\begin{minted}{c}
#include <stdio.h>
int main(void)
{
  int array[5] = {1, 2, 3, 4, 5};
  int index = 0, sum = 0;
  
  while (index <= 10) {
    //@ assert Eva: index_bound: index < 5;
    sum += array[index];
    sum *= 2;
    index ++;
  }
  printf("Sum of array: %d", sum);
  return 0;
}
\end{minted}
\end{minipage}
			\caption{analysis result with \texttt{-eva-precision 3}}
			\label{fig:high-precision}
		\end{subfigure}
	\caption{Identifying potential runtime errors in a C program via the abstract interpretation-based static analyzer {\eva}.}
	\label{fig:example-static-analysis}
\end{figure*}

Albeit with the extensive theoretical study of abstract interpretation, the picture is much less clear on its \emph{parameterization} front: It is challenging to find a set of high-precision parameters to achieve low false-positive rates within a given time budget. The main reasons are two-fold: (i) Off-the-shelf static analyzers often provide a wide range of parameters subject to a huge and possibly infinite joint parameter space. For instance, the parameter setting in \cref{fig:parameter-setting-precision3} consists of 13 external parameters that are highly relevant to the accuracy and efficiency of {\eva}; (ii) The process of seeking highly accurate results typically requires multiple trials of parameter setting and analysis, which generates a large amount of intermediate information such as RTE alarms and analysis time. Nevertheless, few static analyzers provide a fully automated approach to guiding the refinement of abstraction strategies based on such information. Therefore, the use of abstract interpretation-based static analysis tools still relies heavily on expert knowledge and experience.

Some advanced static analysis tools attempt to address the above challenges through various methods. K\"astner et al.~\cite{kastnerAbstractInterpretationIndustry2023}
summarize the four most important abstraction mechanisms in Astr{\'e}e and recommend prioritizing the accuracy of related abstract domains, which amounts to narrowing down the parameter space, though Astr{\'e}e currently does not support automatic parameter generation. \textsc{Goblint}~\cite{saanGoblintAutotuningThreadModular2023, saanGoblintAutotuningThreadModular2024} implements a simple, heuristic autotuning method based on syntactical criteria, which can automatically activate or deactivate abstraction techniques before analysis. However, this method only generates an initial analysis configuration once and does not dynamically adapt to refine the parameter configuration. We defer a detailed survey of related work in this thread to \cref{sec:related-work}.

This paper presents {\parf} -- a fully automated framework to adaptively tune the external parameters of abstract interpretation-based static analyzers. {\parf} models various types of parameters as random variables subject to probability distributions over latticed parameter spaces. Within a given time budget, {\parf} identifies a set of highly accurate parameter settings by incrementally refining the probability distributions based on accumulated intermediate results generated via repeatedly sampling and analyzing.
The core components in {\parf} are the \emph{representation of probability distributions} and the \emph{strategy to refine the distributions}, which together guarantee that the refined joint distribution produces a parameter setting under which {\parf} either (i) yields more accurate analysis in expectation (i.e., \emph{incrementality}), or (ii) in case of analysis failure, terminates with a higher probability (i.e., \emph{adaptivity}).

We have implemented {\parf} on top of {\eva} and compared it against the expert refinement strategy (by trying out increasingly higher precisions) and {\eva}'s official configurations over the {\framac} OSCS benchmark. Experimental results show that {\parf} achieves the highest accuracy
on 34/37 (91.9\%) program repositories with exclusively best results on 12/37 (32.4\%) cases. In particular, {\parf} exhibits promising performance for analyzing complex, large-scale real-world programs. \revision{Moreover, we show that {\parf} can be generalized to improve the performance of other static analyzers such as \mopsa~\cite{DBLP:conf/vstte/JournaultMMO19}.}




\paragraph*{\bf Contributions}
Our main contributions are as follows:
\begin{itemize}
    \item We present {\parf}, a new framework for adaptively tuning external parameters of abstract interpretation-based static analyzers. {\parf} is, to the best of our knowledge, the first \emph{fully automated} approach that supports \emph{incremental} refinement of such parameters. The technical novelty of {\parf} lies in the representation of distributions over a latticed parameter space and the incremental refinement strategy.
    \item We implement {\parf} and demonstrate its effectiveness \revision{and generality} on standard benchmarks. We show that {\parf} outperforms state-of-the-art parameter-tuning strategies by discovering parameter settings leading to more accurate analysis, particularly for programs of a large scale. 
\end{itemize}


%% file: body/background.tex
\section{Background}\label{sec:background}

\subsection{Static Analysis via Abstract Interpretation}\label{sec:static-analysis-ai}

\emph{Static analysis} is the process of analyzing a program without executing its source code. The goal of static analysis is to identify and help users eliminate potential \emph{runtime errors} (RTEs) in the program, e.g., division by zero, overflow in integer arithmetic, and invalid memory accesses. Amongst others, \emph{abstract interpretation}~\cite{cousotAbstractInterpretationUnified1977} is an established technique widely used for \emph{sound} static program analysis: Instead of reasoning about potentially large sets of program states and behaviors, it provides a mechanism to soundly approximate the concrete semantics and thereby reasoning about the program against abstract properties over a certain abstract domain. The soundness nature guarantees that an abstract interpretation-based static analyzer is capable of certifying the absence of RTEs in case no RTE \emph{alarm} is emitted. However, an alarm can be \emph{false positive} due to the abstraction, which does not actually incur RTEs and thus may be excluded by conducting more accurate approximations.

\cref{fig:example-static-analysis} depicts an example of identifying potential runtime errors in a C program using the abstract interpretation-based static analyzer {\eva}. Given a source C program to be analyzed as in \cref{fig:source-prog}, {\eva} emits, under a low-level precision, three alarms signifying potential runtime errors including signed overflow and out-of-bound array index (see \cref{fig:low-precision}). These alarms are expressed as \emph{assertions} written in the ANSI/ISO C specification language (ACSL)~\cite{eva_user_manual}. Nonetheless, by increasing the analysis precision to level 3, {\eva} suffices to rule out the two false-positive alarms concerning signed overflow (see \cref{fig:high-precision}). This is because with \texttt{-eva-precision 3}, {\eva} unrolls the while loop and calculate the value of the variable \texttt{sum} under the assertion \texttt{index < 5}, ensuring that no numerical overflow occurs.

To facilitate the effective use of the tool, {\eva} provides various built-in precision levels, ranging from \verb|-eva-precision 0| to \verb|-eva-precision 11|, each of which packs a group of \emph{fixed} parameter valuations; see \cref{fig:parameter-setting-precision3} for an example. To balance accuracy and efficiency, the commonly adopted \emph{expert refinement strategy} is to tune the parameters by applying {\eva} with increasingly higher precision levels. Such strategy is simple and effective in many cases, yet often yields suboptimal results due to the lack of flexibility in tuning individual parameters; see \cref{sec:experiments}.




\subsection{Parameterization of Static Analysis}
\label{sec:parameterization-of-static-analysis}

\emph{Parameterization} is a typical design approach to enhancing the flexibility and applicability of static analysis tools.
For instance, the parameterization of {\eva} involves \emph{correctness parameters} and \emph{performance tuning} parameters. 
Misusing the former may lead to unsound analysis results and therefore end-users can primarily configure the latter to balance the accuracy and efficiency of abstract interpretation-based static analysis. Throughout the rest of this paper, we consider only performance tuning parameters.



\cref{fig:parameter-setting-precision3} lists a typical parameter setting under \verb|-eva-precision 3| in {\eva}. This setting consists of parameter valuations that are highly relevant to the accuracy and efficiency of abstract interpretation. For instance, users can set the upper bound on the number of times that {\eva} automatically unrolls a loop by configuring the value of \verb|-eva-auto-loop-unroll|, thereby preventing performance degradation caused by excessive loop unrolling; By configuring the value of \verb|-eva-domains|, users can specify a dedicated set of abstract domains used in the analysis. More concretely, the {\eva} command 
\begin{verbatim}
   frama-c *.c -eva -eva-auto-loop-unroll 4
               -eva-domains cvalues,octagon,gauges    
\end{verbatim}
performs abstract interpretation-based static analysis over all C program files in the current directory with loop unrolling limit 4 and abstract domains \verb|cvalues|, \verb|octagon|, and \verb|gauges|. The latter two abstract domains are commonly used to infer different forms of relations between program variables, which both rely on the \verb|cvalues| domain. More detailed parameterization of {\eva} can be found in the manual~\cite{eva_user_manual}.

\subsection{The Complete Lattice Structure}\label{sec:complete-lattice}

Complete lattices are an important mathematical tool used in formalizing the theory of abstract interpretation, as well as in structuring the parameter spaces in our approach. A \emph{complete lattice} \((L, \sqsubseteq)\) consists of a (possibly infinite) carrier set \(L\) and a partial order \(\sqsubseteq\) over \(L\), where every subset \(S \subseteq L\) has both a \emph{greatest lower bound} \(\bigsqcap S \in L\) (also known as the \emph{meet} of \(S\)) and a \textit{least upper bound} \(\bigsqcup S\in L\) (also known as the \emph{join} of \(S\)). For just two elements \(\{x, y\} \subseteq L\), we denote their meet by \(x\sqcap y\) and their join by \(x\sqcup y\). Moreover, we denote by \(\top \defeq \bigsqcup L\) as the \textit{greatest} element of the lattice, and by \(\bot \defeq \bigsqcap \emptyset\) as the \emph{least} element.

Given two complete lattices \((L, \sqsubseteq)\) and \((L', \sqsubseteq')\), a function \(f\colon L \to L'\) is \emph{monotonic} if and only if it respects the partial orders, i.e., for any \(x, y \in L\), \(x \sqsubseteq y\) implies \(f(x)\sqsubseteq' f(y)\) (monotonically increasing) or \(f(y)\sqsubseteq' f(x)\) (monotonically decreasing).



%% file: body/problem_statement.tex
\section{Problem Formulation}\label{sec:problem}

This section formalizes our parameter refining problem. To this end, we first model parameter spaces as complete lattices and then encode static analyzers as monotonic functions over these lattices.


\subsection{Parameter Spaces}
\label{sec:parameter-space}

Given a finite sequence of parameters $P = (P^1, P^2, \ldots, P^n)$, we associate each parameter $P^i$ with its corresponding \emph{parameter space} $PS^i$, which is the (possibly infinite) set of all possible values of parameter $P^i$.
As shown in \cref{fig:parameter-setting-precision3}, commonly used parameters can be classified into four types: (non-negative) integer, Boolean, string, and set-of-strings. For instance, the parameter spaces of the four parameters \verb|-eva-slevel| (integer), \verb|-eva-octagon-through-calls| (Boolean), \verb|-eva-equality-through-calls| (string), and \verb|-eva-domains| (set-of-strings) in {\eva} are as follows:
\begin{align*}    
PS^\text{slevel} &\eeq \PosIntsInf~,\\
PS^\text{octagon-through-calls} &\eeq \{\bFALSE, \bTRUE\}~,\\
PS^\text{equality-through-calls} &\eeq \{\text{`none'}, \text{`formals'}\}~,\\
PS^\text{domains} &\eeq \mathcal{P}\{d_1, d_2, d_3, d_4, d_5\}~.
\end{align*}%
Here, $\PosIntsInf \defeq \Nats \cup \{\infty\}$ is the set of non-negative integers extended with $\infty$. Note that $\infty$ is not a valid parameter value for the underlying static analyzer; rather, it is a symbolic element used to enforce a complete lattice structure of the parameter space (see below). $PS^\text{domains}$ is the power set of the set of five commonly used abstract domains, namely,
\[
    \{\verb|cvalues|, \verb|octagon|, \verb|equality|, \verb|gauges|, \verb|symbolic-locations|\}~.
\]

\paragraph*{\bf Latticed Parameter Spaces}
We observe that every parameter space forms a complete lattice $(PS^i, \sqsubseteq)$:
\begin{itemize}
    \item For an integer parameter, the partial order $\sqsubseteq$ coincides with $\leq$ over $\PosIntsInf$ \revision{(where $k \leq \infty$ for any $k \in \PosIntsInf$)}; The operators $\sqcap$ and $\sqcup$ are then equivalent to $\min$ and $\max$, respectively.
    \item For a Boolean parameter, the partial order $\sqsubseteq$ coincides with implication ${\Rightarrow}$ over the Boolean domain, e.g., $\bFALSE \Rightarrow \bTRUE$; The operators $\sqcap$ and $\sqcup$ are then equivalent to logical connectives $\land$ and $\lor$, respectively.
    \item Since all string parameters in our setting have only two possible values, 
    we map string parameters to the Boolean domain and treat them as Boolean parameters.\revision{\footnote{\revision{For string-typed parameters with $k>2$ possible values: If the parameter exhibits a total order in terms of precision as per \cref{eq:monotonicity}, its space can be mapped to $\{0,1,…,k-1\}$ and thus be treated as an integer-typed parameter. See example in \cref{sec:experiments-generality}.}
    }}
    \item For a set-of-strings parameter, $\sqsubseteq$ coincides with $\subseteq$; The operators $\sqcap$ and $\sqcup$ are then equivalent to set operations $\cap$ and $\cup$, respectively.
\end{itemize}

Next, we define the joint \emph{space of parameter settings} as
\begin{equation*}
    PS \ddefeq PS^1 \times PS^2 \times \cdots \times PS^n~.
\end{equation*}%
Notice that $PS$ also forms a complete lattice $(PS,\preceq)$, where $\preceq$ is the point-wise lifting of $\sqsubseteq$ over individual parameters. Moreover, the meet $\curlywedge$ and join $\curlyvee$ operators on $(PS,\preceq)$ are
\begin{align*}
    p &\curlywedge q \ddefeq (p^1 \sqcap q^1, p^2 \sqcap q^2, \ldots, p^n \sqcap q^n)~,\\
    p &\curlyvee q \ddefeq (p^1 \sqcup q^1, p^2 \sqcup q^2, \ldots, p^n \sqcup q^n)
\end{align*}%
for any parameter settings $p, q \in PS$ with $p = (p^1, p^2,\ldots, p^n)$ and $q = (q^1, q^2, \ldots, q^n)$. For ease of presentation, we abuse the notations $\sqsubseteq$, and $\sqcap, \sqcup$ to denote the partial order $\preceq$ and operators $\curlywedge, \curlyvee$, respectively, in the joint space as well.




\subsection{The Static Analyzer}
\label{sec:static-analyzer}

Next, we show how a static analyzer can be formulated as a monotonic function over the latticed joint parameter space $PS$. To this end, we abstract the procedure of executing static analysis on some program $\textit{prog}$ with some parameter setting $p \in PS$ as a \emph{function} receiving these two parameters and returning a set of (RTE) alarms:
\begin{align*}
  \textit{Analyze}\colon\quad \textit{Prog} \times PS \tto \mathcal P\!\left(A_\text{uni}\right), \ \ (\textit{prog}, p) \mapsto A_p
\end{align*}%
where $\textit{Prog}$ denotes the set of all valid source programs, $A_\text{uni}$ denotes the universe of all possible alarms that can be emitted by the analyzer (which is determined by running the analyzer with the \emph{least} precise parameter setting), and $A_p \subseteq A_\text{uni}$ denotes the set of all alarms emitted under parameter setting $p$.

\paragraph*{\bf Static Analyzer as Monotonic Function}
Note that the codomain of the function $\textit{Analyze}$, i.e., $\mathcal P(A_\text{uni})$, is naturally equipped with a complete lattice structure $(\mathcal P(A_\text{uni}), \subseteq)$. Our incremental refining framework requires that the underlying static analyzer exhibits \emph{monotonicity} over the parameters, that is, for all source programs $\textit{prog} \in \textit{Prog}$ and pairs of parameter settings $p_1, p_2 \in PS$,
%
\begin{align}\label{eq:monotonicity}
   p_1 \ssqsubseteq p_2 \qiimplies \textit{Analyze}(\textit{prog}, p_2) \ssubseteq \textit{Analyze}(\textit{prog}, p_1)
\end{align}%
namely, a \emph{greater} parameter setting (in the latticed joint parameter space) induces \emph{fewer} alarms and thereby \emph{more accurate} analysis. Note that monotonicity is a reasonable and commonly adopted assumption in most state-of-the-art static analyzers~\cite{DBLP:journals/pacmpl/ZhangSZ24};\footnote{{\eva} exhibits monotonicity on most target programs; a corner case is given in~\cite[Section 6.7]{eva_user_manual}, where adding a new domain may unpredictably induce new alarms.} it is also the rationale behind the aforementioned expert refinement strategy where increasing the precision level yields more accurate analysis.



The complete lattice structure of the joint parameter space and the monotonicity of static analyzers allow us to \emph{compare} different parameter settings in terms of accuracy (within a given time budget), which forms the basis of our \emph{parameter refining problem}:
%
%
\begin{framed}
\noindent
\textbf{Problem Statement.} Given a source program $\textit{prog} \in \textit{Prog}$, a time budget $T \in \StrictPosReals$, an abstraction interpretation-based static analyzer $\textit{Analyze}$, and the joint space of parameter settings $PS$ of $\textit{Analyze}$, find a parameter setting $p \in PS$ such that $\textit{Analyze}(\textit{prog}, p)$ returns as few alarms as possible within $T$.
\end{framed}
\noindent
We remark that finding the optimum parameter setting -- which amounts to a brute-force search over a possibly infinite joint parameter space -- is often intractable in practice. We thus aim to develop a fully automated framework to derive parameter settings that yield more accurate analysis than those by the expert refinement strategy. Our framework treats the underlying static analyzer as a black-box function, and thus can be integrated with any abstraction interpretation-based static analyzer (see \cref{sec:Methodology}). For clarity, we take the open-source C analyzer {\eva} as an example; \revision{its extension to other static analyzers is investigated in \cref{sec:experiments}.}

%% file: body/methodology.tex
\begin{figure*}
  \centering
  \includegraphics[width=0.9\linewidth]{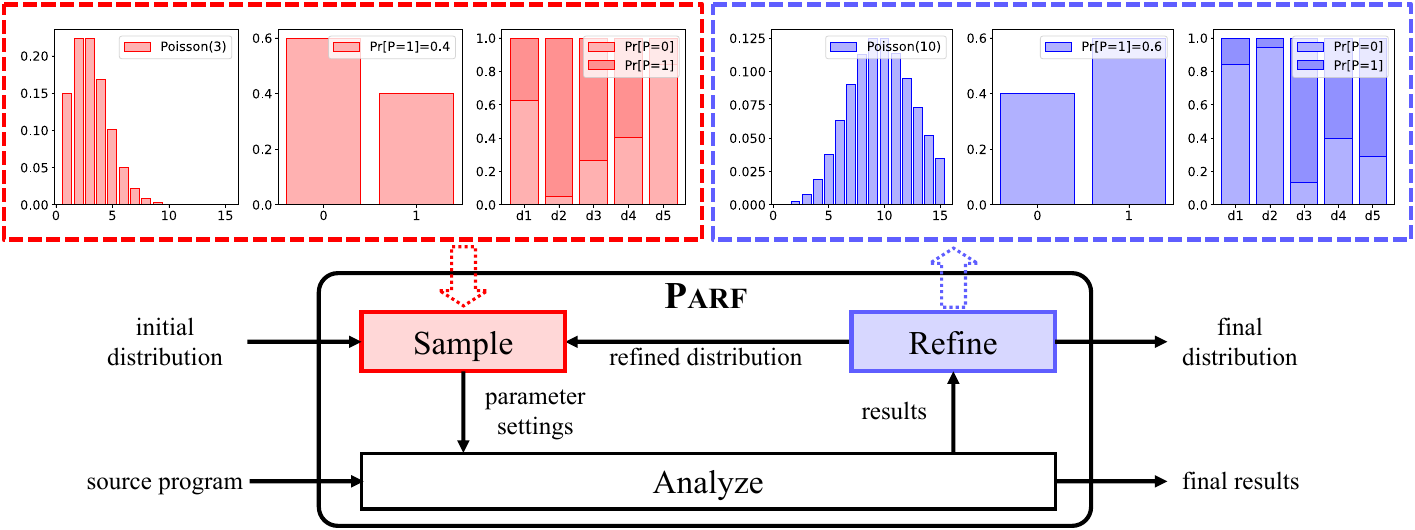}
  \caption{Architecture of the \parf framework. \parf adopts a multi-round iterative mechanism: In each iteration, \parf (i) repeatedly \emph{samples} parameter settings based on the initial or refined probability distribution of parameters, then (ii) uses these parameter settings as inputs to the static analyzer to \emph{analyze} the program, and finally (iii) utilizes the analysis results to \emph{refine} the probability distribution of parameters. \parf continues this process until the prescribed time budget is exhausted, upon which it returns the analysis results of the final round together with the final probability distribution of parameters.}
  \Description{...}
  \label{fig:parf-overview}
\end{figure*}
\begin{figure}
    \centering
    \includegraphics[width=\linewidth]{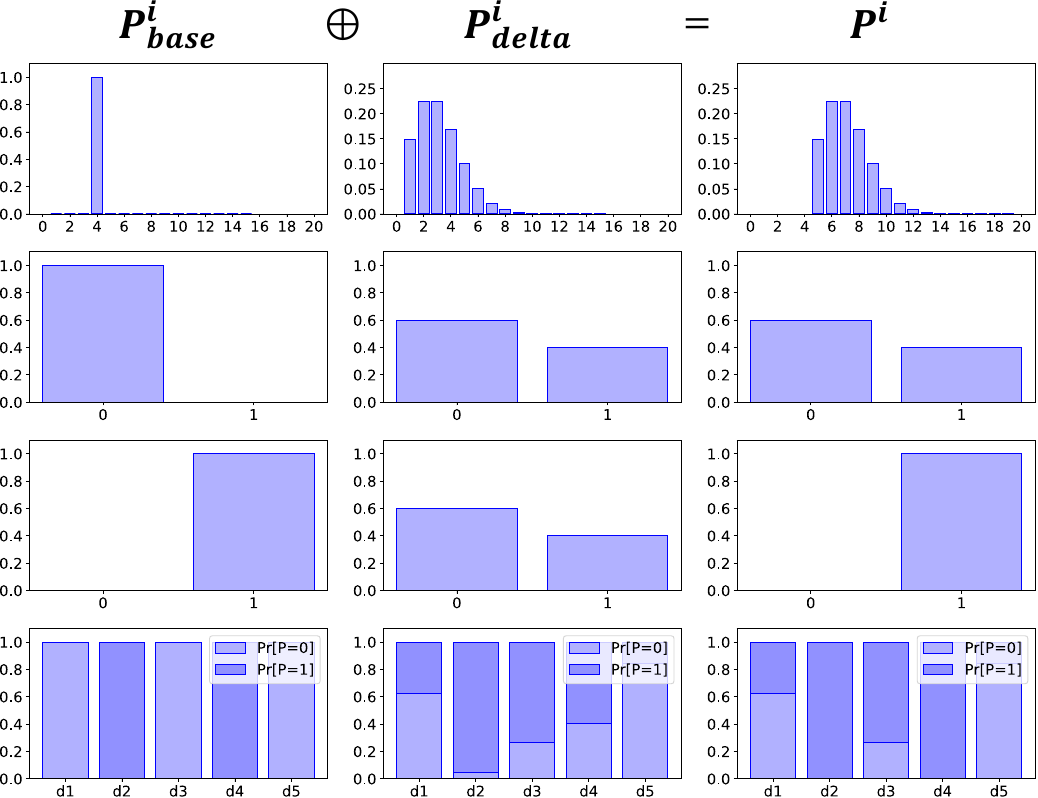}
    \caption{Constructing $P^i$ from $\randomvariable{P}{i}{base}$ and $\randomvariable{P}{i}{delta}$ via $\oplus$. The three columns, from left to right, are $\randomvariable{P}{i}{base}$, $\randomvariable{P}{i}{delta}$, and $P^i$, respectively; The four rows, from top to bottom, correspond to an integer parameter, a Boolean parameter with $Pr[\randomvariable{P}{i}{base}=0]=1$, a Boolean parameter with $Pr[\randomvariable{P}{i}{base}=1]=1$, and a set-of-strings parameter $P^\textnormal{domains}$ over $\mathcal{P}\{d_1, d_2, d_3, d_4, d_5\}$, respectively.}
    \label{fig:two-parts-parameters}
\end{figure}

\section{The Parameter Refining Framework}\label{sec:Methodology}

This section presents our {\parf} framework for Adaptive Parameter ReFining. \cref{fig:parf-overview} illustrates the architecture of {\parf}: It models external parameters of a static analyzer as \emph{random variables} subject to probability distributions over latticed parameter spaces. It \emph{incrementally} refines the probability distributions based on accumulated intermediate results generated by repeatedly sampling and analyzing, thereby ultimately yielding a set of highly accurate parameter settings within a given time budget.


The key components in {\parf} are the \emph{representation of probability distributions} and the \emph{strategy to refine the distributions}; they two cooperate to guarantee that the refined joint distribution produces a parameter setting under which {\parf} either (i) yields more accurate analysis in expectation (i.e., \emph{incrementality}), or (ii) in case of analysis failure, terminates with a higher probability (i.e., \emph{adaptivity}).

\subsection{Probability Distributions of Parameters}

Consider the joint parameter space $PS$ spanned by a fixed set of parameters $\{P^1, P^2, \ldots, P^n\}$.
%
%
We represent every parameter $P^i$ as a \emph{random variable} with sample space $PS^i$. More concretely, $P^i$ is a (measurable) function of the form
\begin{itemize}
    \item $P^i\colon PS^i \to \PosIntsInf$ for an integer parameter,
    \item $P^i\colon PS^i \to \{0, 1\}$ for a Boolean parameter, and
    \item $P^i\colon PS^i \to \{0,1\}^c$, i.e., a $c$-dimensional random vector, for a set-of-strings parameter with \revision{cardinality $c$. Here, $c$ is the number of available strings in the set.}
\end{itemize}
For {\eva}, the only considered set-of-strings parameter is \verb|-eva-domains|, which represents the employed abstract domains.


The (ordered) sequence of parameters is then an $n$-dimensional random vector $P = (P^1, P^2, ..., P^n)$ with sample space $PS$:
\begin{multline*}
    P\colon \quad PS \tto \underbrace{\PosIntsInf\times\cdots\times\PosIntsInf}_{\mathclap{\text{integer parameters}}}\,\ \times\ \underbrace{\{0, 1\}\times\cdots\times\{0, 1\}\vphantom{\PosIntsInf}}_{\mathclap{\text{Boolean parameters}}}\ \times\ \{0,1\}^c
\end{multline*}%
where $c$ is the cardinality of the unique set-of-strings parameter \verb|-eva-domains|. \revision{For cases with $k>1$ set-of-strings parameters, the codomain of $P$ is naturally extended by $\times \{0,1\}^{c_1} \times \cdots \times \{0,1\}^{c_{k-1}}$.}

\paragraph*{\bf Refinable Probability Distribution}
As shown in \cref{fig:parf-overview}, \parf adopts an iterative \procedure{Sample}-\procedure{Analyze}-\procedure{Refine} mechanism to achieve a high-precision (joint) probability distribution of parameter settings within a given time budget. Therefore, the underlying probability distribution needs to feature two abilities: (i) it can effectively \emph{retain} the accumulated knowledge during the iterative procedure, and (ii) it can efficiently \emph{explore} the uncharted parameter space.

To this end, we represent every random variable $P^i$ of the $i$-th parameter as a combination of a \emph{base} random variable and a \emph{delta} random variable:
\begin{equation}\label{eq:variable-combination}
    P^i \qeq \underbrace{\randomvariable{P}{i}{base}}_{\mathclap{\text{for retaining}}} \qoplus \underbrace{\randomvariable{P}{i}{delta}}_{\mathclap{\text{for exploring}}}
\end{equation}%
where $\randomvariable{P}{i}{base}$ is be dedicated to \emph{retaining} the accumulated knowledge whilst $\randomvariable{P}{i}{delta}$ is used to \emph{explore} the parameter space; they share the same sample space and range with $P^i$. $\randomvariable{P}{i}{base}$ follows a one-point distribution, i.e., $Pr[\randomvariable{P}{i}{base} = p^i] = 1$ for some sample $p^i \in PS^i$; The distribution of $\randomvariable{P}{i}{delta}$ depends on the parameter type:
\begin{itemize}
    \item For an integer parameter, $\randomvariable{P}{i}{delta} \sim \textit{Poisson}(\lambda)$ with $\lambda \in \StrictPosReals$.
    \item For a Boolean parameter, $\randomvariable{P}{i}{delta} \sim \textit{Bernoulli}(q)$; $q \in [0, 1]$.
    \item For a set-of-strings parameter with cardinality $c$,  $\randomvariable{P}{i}{delta}$ follows a $c$-dimensional independent joint Bernoulli distribution: $\randomvariable{P}{i}{delta} \sim \textit{Bernoulli}(q_1) \times \cdots \times \textit{Bernoulli}(q_c)$.
\end{itemize}
We now illustrate by \cref{fig:two-parts-parameters} how to construct $P^i$ from $\randomvariable{P}{i}{base}$ and $\randomvariable{P}{i}{delta}$ via the binary operator $\oplus$ as in \cref{eq:variable-combination}:
\begin{itemize}
    \item For an integer parameter, $\oplus$ is equivalent to $+$: Suppose $Pr[\randomvariable{P}{i}{base} = p^i] = 1$ for some non-negative integer $p^i\in \PosIntsInf$,
    $P^i = \randomvariable{P}{i}{base} \oplus \randomvariable{P}{i}{delta}$ is then a rightward shift of $\randomvariable{P}{i}{delta}$ by $p^i$.
    \item For a Boolean parameter, $\oplus$ simulates logical disjunction $\vee$: If $Pr[\randomvariable{P}{i}{base} = 0] = 1$, then $P^i = \randomvariable{P}{i}{delta}$; otherwise if $Pr[\randomvariable{P}{i}{base} = 1] = 1$, then $P^i = \randomvariable{P}{i}{base}$.
    \item For a set-of-strings parameter with cardinality $c$, $\oplus$ is the point-wise lifting of $\vee$ to $c$-dimensional random vectors; see \cref{fig:two-parts-parameters} for an example of $P^\text{domains}$.
\end{itemize}
We can now lift the binary operator $\oplus$ to random vectors as
\begin{align*}
    P \eeq \randomvector{P}{base}\oplus \randomvector{P}{delta} \ddefeq \left(\randomvariable{P}{1}{base} \oplus \randomvariable{P}{1}{delta},\ldots, \randomvariable{P}{n}{base} \oplus \randomvariable{P}{n}{delta}\right)~.
\end{align*}%
Throughout the rest of this section, we will show how to refine $\randomvector{P}{base}$ and $\randomvector{P}{delta}$ in a way such that the composed random variable vector $P$ ensures incrementality and adaptivity.


\subsection{The \parf Algorithm}

\cref{alg:parf} outlines the workflow of \parf: Lines 1-4 correspond to the initialization of variables and the basic analysis of the source program; Lines 5-10 describe the iterative \procedure{Sample}-\procedure{Analyze}-\procedure{Refine} procedure; Lines 11-12 acquire and return the final distribution.

In Line 1, \parf extracts $\randomvector{P}{base}$ and $\randomvector{P}{delta}$ from the initial joint probability distribution $\randomvector{P}{init}$ specifying the initial parameter settings. Here, $\randomvector{P}{base}$ follows a one-point distribution, corresponding to a unique set of parameters, which is actually the default parameter setting of \eva. In Line 2, we analyze the program using the default parameter setting and record the analysis time $time$ and the universe set of alarms $A_{\text{uni}}$. Note that, due to the monotonicity assumption (see \cref{sec:static-analyzer}), alarms obtained in subsequent analyses using refined parameter settings will be subsets of $A_{\text{uni}}$. In Line 3, we set the initial time budget $T_r$ for every round of the \procedure{Sample}-\procedure{Analyze}-\procedure{Refine} process based on two hyper-parameters $\alpha$ and $\beta$. In Line 4, we initialize a counter $count$.

\begin{algorithm}[t]
	\caption{\parf: Adaptive Parameter Refining} 
    \label{alg:parf}
	\begin{algorithmic}[1]
		\renewcommand{\algorithmicrequire}{\textbf{Input:}}
	    \renewcommand{\algorithmicensure}{\textbf{Output:}}
        \REQUIRE The source program $prog$, initial probability distribution of parameters $\randomvector{P}{init}$, and time budget $T$.
        \ENSURE Final probability distribution of parameters $\randomvector{P}{final}$.
        \STATE $\randomvector{P}{base},\ \randomvector{P}{delta}\leftarrow \procedure{Extract}(\randomvector{P}{init})$\,;
        \STATE $time,\ A_{\text{uni}}\leftarrow$ $\procedure{Analyze}$($prog$, $\randomvector{P}{base}$)\,;
        \STATE $T_r\leftarrow$ Max($time \times \alpha$, $T\times \beta$)\,;
        \STATE $count\leftarrow 0$\,;
        \REPEAT
            \STATE $p\_list\leftarrow$ $\procedure{Sample}$($\randomvector{P}{base}\oplus \randomvector{P}{delta},\ num_{\text{sample}}$)\,;
            \STATE $R\_list\leftarrow$ $\procedure{MapAnalyze}$($prog,\ p\_list,\ T_r$)\,;
            \STATE $\randomvector{P}{base},\randomvector{P}{delta}\leftarrow$ $\procedure{Refine}$($p\_list,R\_list,A_{\text{uni}},\randomvector{P}{base},\randomvector{P}{delta}$)\,;
            \STATE $T, T_r, count\leftarrow T - T_r, T_r \times 2, count + 1$\,;
        \UNTIL $T \leq 0$ \OR $count = num_{\text{refine}}$\,;
        \STATE $\randomvector{P}{final}\leftarrow \randomvector{P}{base}$\,;
        \RETURN $\randomvector{P}{final}$\,;
	\end{algorithmic} 
\end{algorithm}

Lines 5-10 represent the iterative \procedure{Sample}-\procedure{Analyze}-\procedure{Refine} process in \parf. In Line 6, \parf samples $num_{\text{sample}}$ (a hyper-parameter) times based on the distribution determined by $\randomvector{P}{base}\oplus \randomvector{P}{delta}$ and returns a list $p\_list$ storing all the parameter settings. In Line 7, \parf separately analyzes the program with each parameter setting in $p\_list$ and obtains a list of results $R\_list$. Each element of $R\_list$ is a pair $\langle p, A \rangle$ storing the parameter setting together with its corresponding analysis alarms. The time of \procedure{MapAnalyze} is limited to $T_r$. In Line 8, \parf utilizes the information from $p\_list$, $R\_list$, and $A_{\text{uni}}$ to refine the random vectors $\randomvector{P}{base}$ and $\randomvector{P}{delta}$ for the next round of analysis. In Line 9, the left time budget and counter are updated, and the time budget of the next round is doubled due to the increase in parameter precision. The loop terminates when either the time budget $T$ is exhausted or the number of refinement iterations reaches the hyper-parameter $num_{\text{refine}}$. Then the one-point distributed $\randomvector{P}{base}$ is returned as the final distribution, which corresponds to a unique set of parameters.


\begin{remark}
Although \eva per se does not support parallel analysis, our \parf algorithm allows for \emph{parallelization}. Specifically, in Line 6 of \cref{alg:parf}, \parf generates a list $p\_list$ containing parameter settings obtained through the \procedure{Sample} function in one go; Then, in Line 7, \parf uses each parameter setting in $p\_list$ for static analysis. As these analyses share no data/control-flow dependencies, they can be threaded in parallel across multiple processes.
\qedT
\end{remark}

\begin{algorithm}[t]
	\caption{\procedure{Refine}: Incremental Refining} 
    \label{alg:refine}
	\begin{algorithmic}[1]
		\renewcommand{\algorithmicrequire}{\textbf{Input:}}
	    \renewcommand{\algorithmicensure}{\textbf{Output:}}
        \REQUIRE List of parameter settings $p\_list$, list of results $R\_list$, universe alarms $A_{\text{uni}}$, and $\randomvector{P}{base}, \randomvector{P}{delta}$.
        \ENSURE Refined distributions $\randomvariable{P}{'}{base}$ and $\randomvariable{P}{'}{delta}$.
        \STATE /* Step 1: Refine $\randomvector{P}{base}$ */
        \STATE $\randomvariable{P}{'}{base}\leftarrow \randomvector{P}{base}$\,;
        \FOR{all $a \in A_{\text{uni}}$}
            \STATE $P_a\leftarrow \top$\,;
            \FOR{all $\langle p,A\rangle \in R\_list$ \AND $a \notin A$}
                \STATE $p_a\leftarrow p_a \sqcap p$\,;
            \ENDFOR
            \IF{$p_a \neq \top$}
                \STATE $\randomvariable{P}{'}{base}\leftarrow \randomvariable{P}{'}{base}\sqcup p_a$\,;
            \ENDIF
        \ENDFOR
        \STATE
        \STATE /* Step 2: Refine $\randomvector{P}{delta}$ */
        \STATE $\eta_{\text{scale}} \leftarrow \frac{2\times |R\_list|+1}{|p\_list|}$\,;
        \STATE $\randomvariable{P}{'}{delta}\leftarrow \eta_{\text{scale}}\otimes \randomvector{P}{delta}$\,;
        
        \RETURN $\randomvector{P}{base}, \randomvector{P}{delta}$\,;
	\end{algorithmic} 
\end{algorithm}

\subsection{The Refinement Strategy}\label{sec:refine}

\cref{alg:refine} describes our incremental strategy to refine the probability distribution of parameter setting, i.e., $\randomvector{P}{base}\oplus \randomvector{P}{delta}$, based on the list of parameter settings $p\_{list}$, list of results $R\_{list}$, and universe set of alarms $A_{\text{uni}}$. The refinement strategies for $\randomvector{P}{base}$ (Step 1, Lines 2-11) and $\randomvector{P}{delta}$ (Step 2, Lines 14-15) are detailed below.

\paragraph*{\bf Refine $\randomvector{P}{base}$}
The refining method for $\randomvector{P}{base}$ involves two nested loops: In the inner loop (Lines 5-7), \parf calculates the \enquote{parameter setting with \emph{lowest} precision} $p_a$ that can eliminate the given false alarm $a$; In the outer loop (Lines 3-11), \parf calculates the \enquote{parameter setting with lowest precision} $\randomvariable{P}{'}{base}$ that can eliminate \emph{all} newly discovered false alarms in this \procedure{Sample}-\procedure{Analyze}-\procedure{Refine} round. The core idea of the refining method for $\randomvector{P}{base}$ can be formalized as
\begin{equation}\label{eq:refine-P-base}
    \randomvariable{P}{'}{base} \ \,\eeq \bigsqcup_{a\in A_{\text{uni}}} p_a \ \,\eeq \bigsqcup_{a\in A_{\text{uni}}}\Bigl( \bigsqcap_{\substack{\langle p,A \rangle\in R\_list \\ a\notin A}} p\Big)~.
\end{equation}%
Here, 
$\bigsqcap\limits_{\substack{\langle p, A \rangle\in R\_list,\, a\notin A}} p$ represents the \emph{greatest lower bound} (for the \emph{lowest} precision) of all sampled parameter settings $p$ which can eliminate false alarm $a$ (signified by $a\notin A$);
$\bigsqcup\limits_{a\in A_{\text{uni}}}p_a$ represents the \emph{least upper bound} (for eliminating \emph{all} false alarms) of all such $p_a$.

\begin{remark}
\revision{Our assumption on the monotonicity of the underlying analyzer is crucial to guarantee the improvement of precision using our refinement strategy: Without monotonicity, the (outer) least upper bound in \cref{eq:refine-P-base} may introduce again a false alarm ruled out by the (inner) greatest lower bound.}
\qedT
\end{remark}

\begin{figure}[t]
    \centering
    \includegraphics[width=\linewidth]{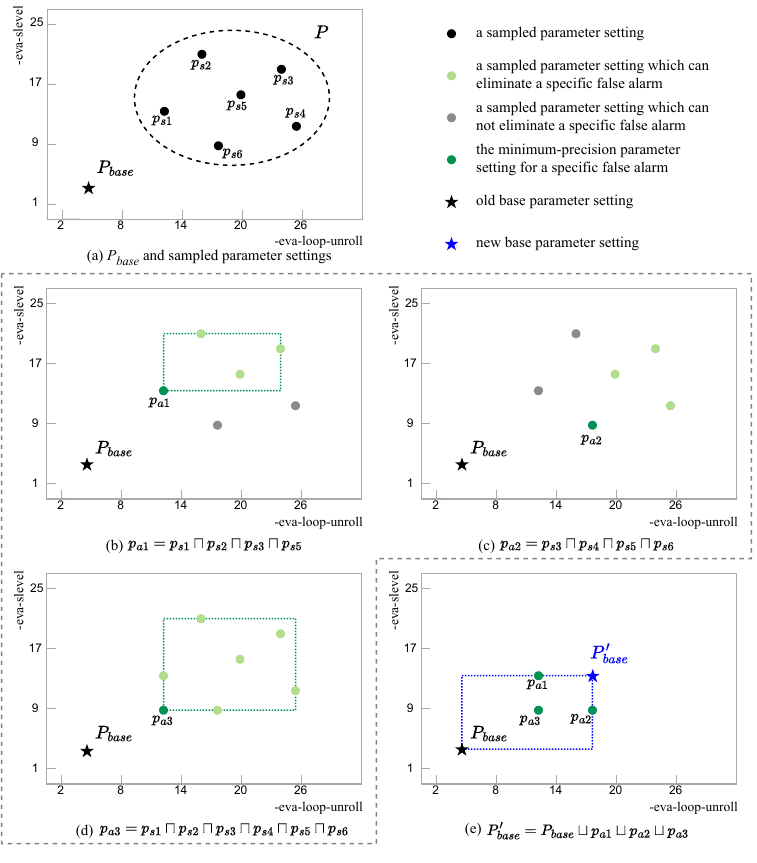}
    \caption{Incremental refinement of $\randomvector{P}{base}$.}
    \Description{...}
    \label{fig:refine-P-base}
\end{figure}

We demonstrate the refinement of $\randomvector{P}{base}$ by means of the following example and \cref{fig:refine-P-base}.

\begin{example}[Refinement of $\randomvector{P}{base}$]
Consider two integer parameters \verb|-eva-slevel| and \verb|-eva-loop-unroll| encoded by random vector $P$ in a two-dimensional parameter space $PS = PS^{\text{slevel}} \times PS^{\text{loop-unroll}}$ as shown in \cref{fig:refine-P-base}(a)-(e). Suppose we have $\randomvector{P}{base}=(4,4)$, the universe set of alarms $A_{\text{uni}}=\{a_1,a_2,a_3\}$, the list of sampled parameter settings $p\_list=(p_{s1},p_{s2},p_{s3},p_{s4},p_{s5},p_{s6})$ (see \cref{fig:refine-P-base}(a)), and the list of analysis results:
\begin{align*}
    R\_list = (&\langle p_{s1},A_1\rangle,\langle p_{s2},A_2\rangle,\langle p_{s3},A_3\rangle,\langle p_{s4},A_4\rangle,\langle p_{s5},A_5\rangle,\langle p_{s6},A_6\rangle) \\
= (&\langle(12, 14), \{a_2\}\rangle, \langle(16, 21), \{a_2\}\rangle, \langle(24, 19), \emptyset\rangle, \\
 &\langle(26, 12), \{a_1\}\rangle, \langle(20, 16), \emptyset\rangle, \langle(18, 9), \{a_1\}\rangle)
\end{align*}%
i.e., false alarm $a_1$ can be eliminated by analyses of $\{p_{s1},p_{s2},p_{s3},p_{s5}\}$, $a_2$ by $\{p_{s3},p_{s4},p_{s5},p_{s6}\}$, and $a_3$ by $\{p_{s1},p_{s2},p_{s3},p_{s4},p_{s5},p_{s6}\}$. Then \cref{fig:refine-P-base}(b)-(d) visualizes the computation of $p_{a1}$, $p_{a2}$, and $p_{a3}$ (Lines 5-7 of \cref{alg:refine}):
\begin{align*}
    p_{a1} &\eeq p_{s1}\sqcap p_{s2}\sqcap p_{s3}\sqcap p_{s5} \eeq (12,14)~,\\
p_{a2} &\eeq p_{s3}\sqcap p_{s4}\sqcap p_{s5}\sqcap p_{s6} \eeq (18,9)~, \\
p_{a3} &\eeq p_{s1}\sqcap p_{s2}\sqcap p_{s3}\sqcap p_{s4} \sqcap p_{s5}\sqcap p_{s6} \eeq (12,9)~.
\end{align*}%
\cref{fig:refine-P-base}(e) depicts the computation of $\randomvariable{P}{'}{base}$ (Lines 3-11 of \cref{alg:refine}):
\begin{align*}
\randomvariable{P}{'}{base} \!\eeq \randomvector{P}{base}\sqcup p_{a1}\sqcup p_{a2}\sqcup p_{a3} \eeq (18,14)
\end{align*}%
$\randomvariable{P}{'}{base}$ thus is the \emph{least precise} parameter setting that can eliminate \emph{all} newly discovered false alarms in the current iteration.
\qedT
\end{example}

Note that, according to the definition of $\sqcup$, $\randomvector{P}{base} \sqsubseteq P'_{\text{base}}$ always holds. This means that $\randomvector{P}{base}$ is incrementally refined.

\paragraph*{\bf Refine $\randomvector{P}{delta}$}
We use the number of (successfully) completed analyses, e.g., $|R\_list|$, and the number of generated parameter settings, e.g., $|p\_list| = num_{\text{sample}}$, to refine $\randomvector{P}{delta}$ in each round, yielding a scaling factor $\eta_{\text{scale}}$ as in Line 14 of \cref{alg:refine}. 
A \emph{larger} value of $\eta_{\text{scale}}$ indicates that more analyses have been completed within time $T_r$, suggesting that a \emph{more extensive exploration} of the parameter space (by scaling up $\randomvector{P}{delta})$ is possible, and vice versa.

More concretely, we define the scaling operator $\otimes$ as
\begin{align}\label{eq:scaling-delta}
    \eta_{\text{scale}}\otimes \randomvector{P}{delta} \eeq \left(\eta_{\text{scale}} \otimes \randomvariable{P}{1}{delta},\cdots,\eta_{\text{scale}} \otimes \randomvariable{P}{n}{delta}\right)~,
\end{align}%
where each component is subject to the distribution type of $\randomvariable{P}{i}{delta}$:
\begin{itemize}
    \item If $\randomvariable{P}{i}{delta}\sim \textit{Poisson}(\lambda)$, then
    \begin{equation*}\label{eq:scale-poisson}
        \eta_{\text{scale}}\otimes \randomvariable{P}{i}{delta} \ssim \textit{Poisson}\left(\eta_{\text{scale}}\times\lambda\right)~.
    \end{equation*}
    \item If $\randomvariable{P}{i}{delta}\sim \textit{Bernoulli}(q)$, then
    \begin{equation*}\label{eq:scale-bernoulli}
        \eta_{\text{scale}}\otimes \randomvariable{P}{i}{delta} \ssim \textit{Bernoulli}\left(1-(1-q)^{\eta_{\text{scale}}}\right)~.
    \end{equation*}
    \item For a set-of-strings parameter with cardinality $c$, $\otimes$ is, again, the point-wise lifting of $\otimes$ to $c$-dimensional random vectors.
\end{itemize}

Observe that the refinement of $\randomvariable{P}{i}{delta}$ as per \cref{eq:scaling-delta} features the following \emph{quantitative} properties:
(i) When $\eta_{\text{scale}}<1$, i.e., more than a half of analyses fail, we have $E[\eta_{\text{scale}}\otimes \randomvariable{P}{i}{delta}]<E[\randomvariable{P}{i}{delta}]$, meaning that the scope of exploration is \enquote{shrunk} in expectation in the next iteration;
(ii) Otherwise if $\eta_{\text{scale}}>1$, we have $E[\eta_{\text{scale}}\otimes \randomvariable{P}{i}{delta}]>E[\randomvariable{P}{i}{delta}]$, meaning that \parf will \enquote{extend} the scope of exploration in expectation.

\paragraph*{\bf Incrementality and Adaptivity.}
By means of separately representing and refining the distributions $\randomvector{P}{base}$ and $\randomvector{P}{delta}$, {\parf} features incrementality and adaptivity. The former guarantees that \enquote{rewardless analyses with low-precision parameters do not occur}, since we have $\randomvector{P}{base} \sqsubseteq P'_{\text{base}}$. The latter allows for an adaptive and quantitative control of the exploration scope, thus avoiding analysis failures while enabling effective search of high-precision parameters.





\begin{remark}
{\parf} is not confined to {\eva}; rather, it can be readily integrated with any static analyzer \revision{exhibiting monotonicity over the parameters}. This is because (i) {\parf} treats the static analyzer as a black box, and (ii) {\parf} covers a wide range of parameter types commonly used in static analyzers. Moreover, we can extend {\parf} without substantial changes to incorporate extra types of parameters, e.g., real-valued parameters, by formulating a latticed parameter space with Gaussian distributions. \revision{As an example, we interface {\parf} with the static analyzer {\mopsa}~\cite{DBLP:conf/vstte/JournaultMMO19} in \cref{sec:experiments}.}

\revision{Moreover, due to its black-box nature, {\parf} exhibits no strong correlation with abstract interpretation and thus can be paired with other static analysis techniques. We opt for abstract interpretation as it is a typical analysis technique featuring monotonicity.}
%
\qedT
\end{remark}

%% file: body/experiments.tex
\section{Implementation and Experiments}\label{sec:experiments}

The experiments aim to answer the following research questions:
\begin{itemize}
    \item[\textbf{RQ1:}] How does \parf compare against other parameter-selecting strategies?
    \item[\revision{\textbf{RQ2:}}] \revision{How does \parf perform on different hyper-parameters?}
    \item[\revision{\textbf{RQ3:}}] \revision{Can \parf be generalized to other static analyzers?}
    \item[\textbf{RQ4:}] Can \parf improve \framac in verification competitions?
\end{itemize}


\revision{To answer \textbf{RQ1}, \textbf{RQ2}, and \textbf{RQ4}, we implement {\parf} as a plugin of \framac, which is an open-source collaborative platform dedicated to scalable source-code analysis of C programs~\cite{DBLP:conf/rv/KosmatovS16}. For \textbf{RQ3}, we further integrate {\parf} with the static analyzer {\mopsa}~\cite{DBLP:conf/vstte/JournaultMMO19}. \parf supports parallelization across multiple processes: We adopt a four-process parallelization mechanism for \textbf{RQ1}, \textbf{RQ2}, and \textbf{RQ3}, and a single process for \textbf{RQ4}.\footnote{\revision{For the latter, parallelization does not make a difference for SV-COMP as it counts the total CPU time for all processes.}} The experiments for \textbf{RQ1}, \textbf{RQ2}, and \textbf{RQ4} are conducted on an 8-core Apple M2 processor with 16GB RAM running 64-bit macOS Sonoma 14; \textbf{RQ3} is evaluated on a 16-core Intel i7 processor with 16GB RAM running Arch Linux (as \mopsa~\cite{DBLP:conf/vstte/JournaultMMO19} runs only on Linux). Experimental results reported in this paper can be found in the Docker image at \url{https://hub.docker.com/r/parfdocker/parf}.}

\begin{table}[t]
    \caption{Initial probability distributions for \eva.}
    \label{tab:initial-probability}
    \centering\small
    \begin{tabular}{lccc}
        \toprule
        $P^i$ & Type  & $\randomvariable{P}{i}{base}$ & $\randomvariable{P}{i}{delta}$ \\
        \midrule
        $P^{\text{min-loop-unroll}}$ & integer & 0 & $\textit{Poisson}(0.4)$ \\
        $P^{\text{auto-loop-unroll}}$ & integer & 0 & $\textit{Poisson}(10)$ \\
        $P^{\text{widening-delay}}$ & integer & 1 & $\textit{Poisson}(0.5)$ \\
        $P^{\text{partition-history}}$ & integer & 0 & $\textit{Poisson}(0.4)$ \\
        $P^{\text{slevel}}$ & integer & 0 & $\textit{Poisson}(20)$ \\
        $P^{\text{ilevel}}$ & integer & 8 & $\textit{Poisson}(2)$ \\
        $P^{\text{plevel}}$ & integer & 10 & $\textit{Poisson}(10)$ \\
        $P^{\text{subdivide-non-linear}}$ & integer & 0 & $\textit{Poisson}(2.5)$ \\
        $P^{\text{remove-redundant-alarms}}$ & Boolean & 0 & $\textit{Bernoulli}(0.5)$ \\
        $P^{\text{split-return}}$ & string & 0 & $\textit{Bernoulli}(0.5)$ \\
        $P^{\text{equality-through-calls}}$ & string & 0 & $\textit{Bernoulli}(0.5)$ \\
        $P^{\text{octagon-through-calls}}$ & Boolean & 0 & $\textit{Bernoulli}(0.5)$ \\
        $P^{\text{domains}}$ & set-of-strings & $(1,0,0,0,0)$ & $\textit{Bernoulli}(0.5)^5$ \\
        \bottomrule
    \end{tabular}
\end{table}

\subsection{Experimental Settings}

\paragraph*{Settings of \textnormal{\parf}}
\parf provides an optional interface for setting the initial parameter distribution, which allows users to provide a prior probability distribution $\randomvector{P}{init}$ based on their own experience. In absence of user-specified $\randomvector{P}{init}$, \parf sets $\randomvector{P}{base}$ and $\randomvector{P}{delta}$ according to \cref{tab:initial-probability}. In our experiments, we use the default initial distribution as in \cref{tab:initial-probability} and set the hyper-parameters $\alpha$, $\beta$, $num_{\text{sample}}$, and $\textit{num}_{\text{refine}}$ in \cref{alg:parf} to $0.1$, $2$, $4$, and $7$, respectively. \revision{We examine the effect of these hyper-parameters in \textbf{RQ2}.}

\paragraph*{Benchmarks}
The first benchmark suite we use is the Frama-C official Open Source Case Study (OSCS) benchmarks\footnote{Available at \url{https://git.frama-c.com/pub/open-source-case-studies}. We filtered out several projects which contain, e.g., configuration issues or test suites.}. This benchmark set includes many real-world C projects, such as the X509 parser project (a {\framac}-verified static analyzer)~\cite{x509-parser,CAV24-LLM4FM}. \cref{tab:experiment-results} lists the detailed information about the benchmarks as described below:
\begin{itemize}
    \item Benchmark name: Name of each benchmark in C.
    \item LOC (lines of code): Size of each benchmark's source files.
    \item \#statements: The number of statements analyzed by \framac in each benchmark. A statement is the smallest syntactic functional element of the target program~\cite{buhlerEVAEvolvedValue2017}.
    \item \verb|-eva-precision|: This index reflects how fast each benchmark can be analyzed. A larger number indicates that the analysis can terminate in less time under the same parameter settings; see details in \cref{sec:parf-selecting-strategies}.
\end{itemize}

The second benchmark suite is collected from the verification tasks of SV-COMP 2022\footnote{Available at \url{https://gitlab.com/sosy-lab/benchmarking/sv-benchmarks}.}~\cite{beyerStaticAnalyzerFramaC2022,beyerProgressSoftwareVerification2022}, where {\framac} participated in the NoOverflows\footnote{Available at \url{https://sv-comp.sosy-lab.org/2022/benchmarks.php}.} category with a specific version called {\framacsv}.


\paragraph*{Baselines and Time Budget}


We compare our approach \parf against three baselines: \default, \official, and \expert, which correspond to three existing parameter-selecting strategies. The \default strategy uses default parameter settings of \eva \revision{or \mopsa} to perform static analysis; The \official strategy uses official parameter settings provided by \framac together with the OSCS benchmarks, which can be considered as \enquote{high quality} parameter settings; \expert is a dynamic parameter-tuning strategy for \eva, which sequentially increases the parameters from \verb|-eva-precision 0| to \verb|-eva-precision 11| for analysis until the given time budget is exhausted or the highest precision level is reached. We set the total time budget for each benchmark to 1 hour.


\begin{figure*}[t]
    \begin{minipage}[b]{\textwidth}
    \begin{table}[H]
      \centering
        \caption{Experimental results in terms of \textbf{RQ1} \revision{and \textbf{RQ3}}.}
        \label{tab:experiment-results}
        \centering\footnotesize
        \begin{tabular}{lrrrrrrrrr}
            \toprule
            \multicolumn{4}{c}{\footnotesize{\textbf{OSCS Benchmark Details}}} & \multicolumn{4}{c}{\footnotesize{\textbf{\#Alarms of \framac} \revision{(\textbf{RQ1})}}} & \multicolumn{2}{c}{\footnotesize{\textbf{\revision{\#Alarms of \mopsa}} \revision{(\textbf{RQ3})}}} \\
            \cmidrule(lr){1-4} \cmidrule(lr){5-8} \cmidrule(lr){9-10}
            Benchmark name & LOC & \#statements & \verb|-eva-precision| & \default & \expert & \official & \parfstrategy & \revision{\default} & \revision{\parfstrategy} \\
            \cmidrule(lr){1-4} \cmidrule(lr){5-8} \cmidrule(lr){9-10}
            2048 & 440 & 329 & 6 & 7 & 5 & 7 & \best{4} & \revision{141}  & \revision{\best{67}}  \\
            chrony & 37177 & 41 & 11 & 9 & \similar{7} & 8 & \similar{7} & \revision{--} & \revision{--} \\
            debie1 & 8972 & 3243 & 2 & 33 & 3 & \best{1} & 19  & \revision{8245} & \revision{\best{5656}} \\
            genann & 1183 & 1042 & 9 & 236 & \similar{69} & 77 & \similar{69} & \revision{\similar{1308}} & \revision{\similar{1308}} \\
            gzip124 & 8166 & 4835 & 1 & \similar{802} & 885 & 866 & \similar{810} & \revision{--} & \revision{--} \\
            hiredis & 7459 & 87 & 11 & 9 & \similar{0} & 9 & \similar{0} & \revision{\similar{43}} & \revision{\similar{43}} \\
            icpc & 1302 & 424 & 11 & 9 & \similar{1} & \similar{1} & \similar{1} & \revision{11} & \revision{\best{10}} \\
            jsmn-ex1 & 1016 & 1219 & 11 & 58 & \similar{1} & \similar{1} & \similar{1} & \revision{1762} & \revision{\best{1253}} \\
            jsmn-ex2 & 1016 & 311 & 11 & 68 & \similar{1} & \similar{1} & \similar{1} & \revision{87} & \revision{\best{86}} \\
            kgflags-ex1 & 1455 & 474 & 11 & 11 & \similar{0} & 11 & \similar{0} & \revision{\similar{280}} & \revision{\similar{280}} \\
            kgflags-ex2 & 1455 & 736 & 10 & 33 & \similar{19} & 33 & \similar{19} & \revision{\similar{386}} & \revision{\similar{386}} \\
            khash & 1016 & 206 & 11 & 14 & \similar{2} & 14 & \similar{2} & \revision{\similar{19}} & \revision{\similar{19}} \\
            kilo & 1276 & 1078 & 2 & 523 & 445 & 688 & \best{429} & \revision{\similar{5299}} & \revision{\similar{5290}} \\
            libspng & 4455 & 2377 & 6 & 186 & 122 & 122 & \best{113} & \revision{--} & \revision{--} \\
            line-following-robot & 6739 & 857 & 11 & \similar{1} & \similar{1} & \similar{1} & \similar{1} & \revision{--} & \revision{--} \\
            microstrain & 51007 & 3216 & 6 & 1177 & 616 & 646 & \best{598}  & \revision{\similar{6237}} & \revision{\similar{6196}} \\
            mini-gmp & 11706 & 628 & 6 & 83 & \similar{71} & 83 & \similar{71} & \revision{513} & \revision{\best{491}} \\
            miniz-ex1 & 10844 & 3659 & 1 & 2291 & \similar{1832} & 2291 & \similar{1828} & \revision{\similar{3020}} & \revision{\similar{3004}} \\
            miniz-ex2 & 10844 & 5589 & 1 & 2742 & 2220 & 2742 & \best{2172} & \revision{\similar{3916}} & \revision{\similar{3899}} \\
            miniz-ex3 & 10844 & 3747 & 1 & 577 & 552 & 577 & \best{442} & \revision{\similar{2808}} & \revision{\similar{2792}} \\
            miniz-ex4 & 10844 & 1246 & 4 & 258 & 217 & 258 & \best{189} & \revision{\similar{162}} & \revision{\similar{162}} \\
            miniz-ex5 & 10844 & 3430 & 1 & 425 & 402 & 425 & \best{377} & \revision{1575} & \revision{\best{1474}} \\
            miniz-ex6 & 10844 & 2073 & 1 & 220 & 198 & 220 & \best{173} & \revision{1197} & \revision{\best{1075}} \\
            monocypher & 25263 & 4126 & 1 & 606 & \similar{570} & \similar{568} & 606 & \revision{TO} & \revision{TO} \\
            papabench & 12254 & 36 & 11 & \similar{1} & \similar{1} & \similar{1} & \similar{1} & \revision{--} & \revision{--} \\
            qlz-ex1 & 1168 & 229 & 11 & 68 & \similar{11} & 68 & \similar{11} & \revision{\similar{82}} & \revision{\similar{82}} \\
            qlz-ex2 & 1168 & 75 & 11 & \similar{8} & \similar{8} & \similar{8} & \similar{8} & \revision{\similar{50}} & \revision{\similar{50}} \\
            qlz-ex3 & 1168 & 294 & 8 & 94 & 82 & 94 & \best{75} & \revision{--} & \revision{--} \\
            qlz-ex4 & 1168 & 164 & 11 & 17 & \similar{13} & 17 & \similar{13} & \revision{--}  & \revision{--} \\
            safestringlib & 29271 & 13029 & 6 & 855 & \best{256} & 300 & 356 & \revision{--} & \revision{--} \\
            semver & 1532 & 728 & 9 & 29 & \similar{22} & 25 & \similar{22} & \revision{3556} & \revision{\best{2850}} \\
            solitaire & 338 & 396 & 11 & 216 & \similar{18} & 213 & \similar{18} & \revision{700} & \revision{\best{663}} \\
            stmr & 781 & 500 & 6 & 63 & \similar{58} & 59 & \similar{58} & \revision{\similar{1391}} & \revision{\similar{1391}} \\
            tsvc & 5610 & 5478 & 4 & 413 & \similar{355} & 379 & \similar{356} & \revision{--} & \revision{--} \\
            tutorials & 325 & 89 & 11 & 5 & 1 & 5 & \best{0} & \revision{--} & \revision{--} \\
            tweetnacl-usable & 1204 & 659 & 11 & 126 & \similar{25} & 30 & \similar{25} & \revision{667} & \revision{\best{657}} \\
            x509-parser & 9457 & 3112 & 3 & 208 & 198 & 198 & \best{187} & \revision{364} & \revision{\best{339}} \\
            \midrule
            \multicolumn{4}{l}{\footnotesize{Overall (\similar{tied-best}+\best{exclusively best})}} & \footnotesize{3/37} & \footnotesize{23/37} & \footnotesize{8/37} & \footnotesize{34/37 (91.9\%)} 
            & \revision{\footnotesize{14/27 (51.9\%)}}
            & \revision{\footnotesize{26/27 (96.3\%)}} \\

            \multicolumn{4}{l}{\footnotesize{Overall (\best{exclusively best})}} & \footnotesize{0/37} & \footnotesize{1/37}  & \footnotesize{1/37} & \footnotesize{12/37 (32.4\%)}
            & \revision{\footnotesize{0/27 (0.0\%)}}
            & \revision{\footnotesize{12/27 (44.4\%)}}  \\
            \bottomrule
        \end{tabular}
    \end{table}
    \end{minipage}
\end{figure*}

\subsection{RQ1: \parf vs. Other Strategies}\label{sec:parf-selecting-strategies}

\cref{tab:experiment-results} shows our experimental results in response of \textbf{RQ1}. Let us first illustrate the evaluation method for analysis results of the four parameter-selection strategies: \default, \expert, \official, and \parf. We mark the result with the least number of alarms and not tied with the other three as the \best{exclusively best}. For instance, the analysis result of the \parf strategy for the first benchmark 2048 has \best{4} alarms. We also mark the results with the same least number of alarms as \similar{tied-best}.  Given that some analysis results contain thousands of alarms, it is unfair and unreasonable to distinguish two results based on minor differences in the number of alarms. Therefore, we mark multiple sets of results with $\leq 1\%$ difference in the number of alarms from the least one as \similar{tied-best}. For instance, the \expert and \parf analysis results for benchmark miniz-ex1 contain \similar{1832} and \similar{1828} alarms as tied-best respectively ($1832-1828 \leq 0.01\times 1832$). \revision{We adopt the same convention for \textbf{RQ3}.}

\cref{tab:experiment-results} shows that {\parf} significantly outperforms other strategies for {\framac} in terms of accuracy on the OSCS dataset: Overall, \parf achieves the best results (i.e., exclusively best or tied-best) on 34/37 (91.9\%) benchmarks, while the \default, \expert, and \official strategies achieve the best results on 3/37, 23/37, and 8/37 benchmarks, respectively. Furthermore, over 12/37 (32.4\%) benchmarks, \parf achieves exclusively best results, while the competitors achieve exclusively best results only on 0 or 1 specific benchmark.

The above results exhibit the capacity of \parf to achieve high-accuracy analysis results in various real-world scenarios, rather than specific scenarios. This is because the OSCS benchmarks possess diversity and representativeness, with project sizes ranging from 338 to 51,007 LOC and numbers of analyzed statements ranging from 36 to 13,029. Moreover, the values in the \texttt{-eva-precision} column intuitively reflect the analysis complexity for a given benchmark, as they are the highest \texttt{-eva-precision} parameter values identified by \expert strategy within 1 hour. Therefore, benchmarks with lower complexity require less analysis time under the same parameters, resulting in higher values in the \texttt{-eva-precision} column within the given time. Meanwhile, the range of \texttt{-eva-precision} also indicates the diversity of the OSCS benchmarks.

Interestingly, among around half (18/37) of the OSCS benchmarks with high \texttt{-eva-precision} values ($\geq 9$), \parf achieves tied-best results in almost all cases, with only one exclusively best (benchmark tutorials). Considering that \parf achieves the exclusively best results on 12/37 (32.4\%) benchmarks, \parf performs exclusively best on 57.9\% of the remaining 19 benchmarks with low-to-medium \texttt{-eva-precision} values ($\leq 8$). The reason behind this observation is as follows. On one hand, benchmarks with high \texttt{-eva-precision} values have low analysis complexity, allowing various strategies to readily find high-accuracy or even the most accurate analysis results within the given time. Therefore, \parf can only achieve tied-best results with them. On the other hand, benchmarks with low-to-medium \texttt{-eva-precision} values have high analysis complexity, making it difficult for \default, \official, and \expert strategies to avoid analysis failure and find the most accurate parameters. However, \parf's adaptivity enables it to handle such challenging tasks. In summary, this demonstrates that \emph{\parf is particularly suitable for analyzing complex, large-scale real-world programs}.

\begin{table}[t]
    \caption{\revision{Average timings of the experiments in terms of \textbf{RQ1}.}}
    \label{tab:running-time}
    \centering\small
    \begin{tabular}{lrrrr}
        \toprule
        \multirow{2}{*}{\revision{\textbf{Phase}}} & \multicolumn{4}{c}{\revision{\textbf{Average Time for \framac (s)}}}\\
        \cmidrule(lr){2-5}
          & \revision{\default} & \revision{\expert} & \revision{\official} & \revision{\parf}  \\
        \midrule
        \revision{Identification of $p$} & \revision{0.00} & \revision{2340.59} & \revision{--} & \revision{\textbf{1315.97}} \\
        \revision{Analysis under $p$} & \revision{9.00}  & \revision{465.05} & \revision{21.78} & \revision{78.92} \\
        \bottomrule
    \end{tabular}
\end{table}

\begin{figure*}[t]
    \begin{tikzpicture}
            \begin{axis}[  
                width=0.95\textwidth,
                height=4.3cm,
                ylabel={\#alarms},  
                xlabel={$(\alpha, \beta)$}, 
                xlabel style={xshift=-0.275\textwidth, yshift=3ex},
                axis y line*=left,  
                tick align=inside,
                xmin=-1, xmax=30, ymin=10.5, ymax=17,
                xtick={1,2,3,4,5,6,7,8,9},
                xticklabels={(0.067,1.33), (0.067,2), (0.067,3), (0.1,1.33), (0.1,2), (0.1,3),(0.15,1.33),(0.15,2),(0.15,3)},
                xticklabel style={rotate=38, anchor=east, yshift=-0.5ex,font=\scriptsize},
                legend style={at={(0.015,0.95)},anchor=north west,font=\footnotesize,inner sep=0.5pt},
            ]  
                \addplot [Maroon, mark=*, mark size=1pt] coordinates {  
                    (1,12.89) (2,13.39) (3,12.89) (4,13.33) (5,13.00) (6,14.67) (7,13.00) (8,13.72) (9,12.89) 
                };\addlegendentry{\#alarms}
                \addplot[blue, mark=square*, mark size=1pt] coordinates {(0,-100)}; \addlegendentry{identif.~time}
  	             \addplot[dashed, color=ForestGreen, line width=0.5pt] coordinates {(0,-100)};\addlegendentry{selected config.}
            \end{axis}  
            \begin{axis}[ 
                width=0.95\textwidth,
                height=4.3cm,
                axis y line*=right,  
                axis x line=none,  
                ylabel={identif.~time (s)},  
                tick align=inside,
                xmin=-1, xmax=30, ymin=0, ymax=1100,
                clip=false,
                axis on top=false,
            ]  
                \addplot [blue, mark=square*, mark size=1pt] coordinates {  
                    (1,209.72) (2,174.22) (3,207.22) (4,182.00) (5,166.39) (6,274.17) (7,248.89) (8,201.44) (9,190.44) 
                };\label{plot_two}
                \draw[dashed, color=ForestGreen, line width=0.5pt] (5,0) -- (5,1100);\label{plot_tree}
                \addplot [blue, mark=square*, mark size=1pt] coordinates {  
                    (12,86.39) (14,166.39) (16,256.06) (18,271.06) (20, 312.94)
                };
                \draw[dashed, color=ForestGreen, line width=0.5pt] (14,0) -- (14,1100);
                \addplot [blue, mark=square*, mark size=1pt] coordinates {  
                    (23,166.39) (24,606.89) (25,767.67) (26,954.33) (27,1011.61) (28, 1001.22)
                };
                \draw[dashed, color=ForestGreen, line width=0.5pt] (27,0) -- (27,1100);
            \end{axis}  

            \begin{axis}[  
                width=0.95\textwidth,
                height=4.3cm,
                ylabel={\#alarms},  
                xlabel={$num_{\text{sample}}$}, 
                xlabel style={xshift=0.041\textwidth},
                axis y line*=left,  
                tick align=inside,
                xmin=-1, xmax=30, ymin=10.5, ymax=17,
                xtick={12, 13,14,15,16,17,18,19,20},
                xticklabels={2,3,4,5,6,7,8,9,10},
            ]  
                \addplot [Maroon, mark=*, mark size=1pt] coordinates {  
                    (12,15.50) (14,13.00) (16,13.00) (18,12.89) (20, 12.89)
                };
                \label{plot_one}
            \end{axis}

            \begin{axis}[  
                width=0.95\textwidth,
                height=4.3cm,
                ylabel={\#alarms},  
                xlabel={$num_{\text{refine}}$}, 
                xlabel style={xshift=0.307\textwidth},
                axis y line*=left,  
                tick align=inside,
                xmin=-1, xmax=30, ymin=10.5, ymax=17,
                xtick={23,24,25,26,27,28},
                xticklabels={3,4,5,6,7,8},
            ]  
                \addplot [Maroon, mark=*, mark size=1pt] coordinates {  
                    (23,13.00) (24,12.94) (25,11.56) (26,11.17) (27,11.06) (28, 11.17)
                };
                \label{plot_one}
            \end{axis} 
            
            \label{fig:subfig3}  
    \end{tikzpicture}
    \begin{tikzpicture}[remember picture, overlay]
        \foreach \i in {0,...,3}{
			\draw[color=white, line width=5pt] (-11.45+0.23*\i,0.9) -- (-11.45+0.23*\i,3.9);
            \draw[color=white, line width=5pt] (-6.01+0.23*\i,0.9) -- (-6.01+0.23*\i,3.9);
        }
	\end{tikzpicture}
    \caption{\revision{Effect of hyper-parameters on the performance of {\parf} in terms of accuracy (\#alarms) and efficiency (time).}}  
    \label{fig:hyper-parameters}
\end{figure*}
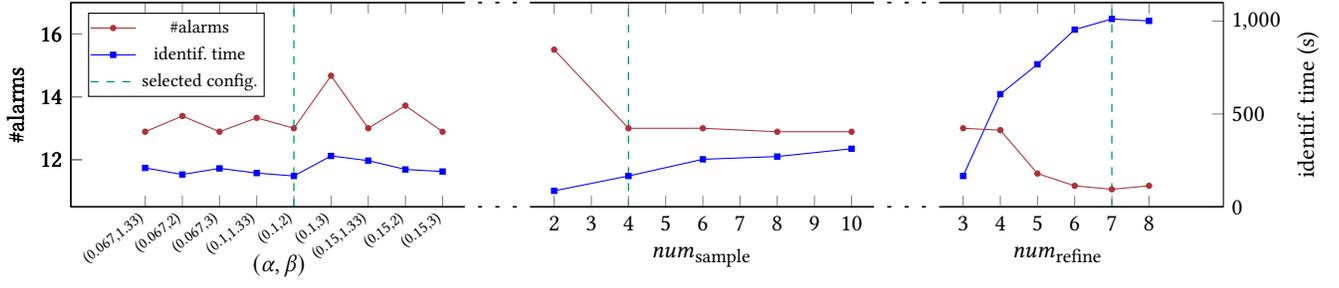

Note that, as the computation in \parf is randomized, we repeatedly conduct two experiments with \parf, limiting the analysis time budget of each benchmark to 30 minutes in each experiment, and select the better analysis result (the same applies to \textbf{RQ3}). This is to simulate the real situation of experts using static analyzers: trying to find the best analysis result within a given total time budget. However, this cannot completely eliminate randomness; \parf still has the possibility to fail to find the best analysis result within the time limit (e.g., benchmarks debie1, monocypher, and safestringlib).

\begin{table}[t]
    \caption{Initial probability distributions for \mopsa.}
    \label{tab:initial-probability-of-mopsa}
    \centering\small
    \begin{tabular}{lccc}
        \toprule
        $P^i$ & Type & $\randomvariable{P}{i}{base}$ & $\randomvariable{P}{i}{delta}$ \\
        \midrule
        $P^{\text{-max-set-size}}$ & integer & 1 & $\textit{Poisson}(1.0)$ \\
        $P^{\text{-loop-unrolling}}$ & integer & 1 & $\textit{Poisson}(2.0)$ \\
        $P^{\text{-widening-delay}}$ & integer & 0 & $\textit{Poisson}(0.5)$ \\
        $P^{\text{-numeric}}$ & string & 0 & $\textit{Poisson}(0.5)$ \\
        $P^{\text{-loop-decr-it}}$ & Boolean & 0 & $\textit{Bernoulli}(0.5)$ \\
        $P^{\text{-loop-no-cache}}$ & Boolean & 0 & $\textit{Bernoulli}(0.5)$ \\
        \bottomrule
    \end{tabular}
\end{table}

\paragraph*{\it \revision{Running Time under Different Strategies}}
\revision{\cref{tab:running-time} shows the timings (averaged over the OSCS benchmarks) of (i) identifying the final parameter setting $p$ and (ii) analyzing the source program under $p$. As formulated in \cref{sec:problem}, the application scenario of our approach is to automatically find a highly accurate parameter setting within a given time budget. In this context, we find it less interesting to compare the \emph{analysis time under the specific parameter setting} produced by different strategies (bottom row of \cref{tab:running-time}), since parameters of higher precision typically require more time to perform static analysis. In contrast, we are concerned with the \emph{time for identifying high-precision parameter settings} (first row of \cref{tab:running-time}), where (i) The \default strategy provides low-precision parameters in no time but suffers from significantly low accuracy; (ii) The \official strategy yields high-precision parameters prepared by {\framac} officially for the OSCS benchmarks, but it does not specify the human effort and time invested in identifying these parameters. This implies that manually selecting parameters is still necessary for programs beyond OSCS; (iii) The \expert strategy represents a simple yet effective automated dynamic parameter-tuning strategy employed by experts, which achieves results with significantly higher accuracy than the \default and \official strategies within a reasonable time budget (1 hour in our case); (iv) \parf identifies the most accurate parameter settings in 91.9\% benchmarks only with around half of the time used by \expert (the only automated, dynamic competitor).}

\begin{remark}
 Our \parf strategy adopts a four-process parallelization mechanism. We remark that a similar mechanism does not apply to the other parameter-selecting strategies: (i) The parameters given by \default and \official are fixed; (ii) For the \expert strategy, parallelizing the trials with different precision levels does often not bring observable enhancement in efficiency, since it is still the trial with the highest precision level that consumes significantly more time than those with lower precision levels.
\qedT
\end{remark}

\subsection{\revision{RQ2: Effect of Hyper-Parameters}}

\revision{\cref{fig:hyper-parameters} depicts the effect of hyper-parameters on {\parf}'s performance -- averaged over 18/37 OSCS benchmarks (with $\texttt{-eva-precision} \geq 9$)\footnote{\revision{This subset of benchmarks features a relatively low analysis complexity and thus our evaluation of \textbf{RQ2} can be completed in a reasonable amount of time.}} -- in terms of accuracy (\#alarms) and efficiency (time for identifying the final parameter setting):
\begin{itemize}
    \item Effect of $\alpha \in (0,1)$ and $\beta > 1$: These two hyper-parameters control the initial time budget $T_r$ for every round of the \procedure{Sample}-\procedure{Analyze}-\procedure{Refine} process in the \parf algorithm. We observe that \parf is not sensitive to the values of $\alpha$ and $\beta$.
    \item Effect of $num_{\text{sample}}$: This parameter upper-bounds the number of samples in one refinement iteration. Observe that both the analysis accuracy and the time remain stable for $num_{\text{sample}}\geq 6$ (due to the time budget $T_r$ per iteration). In our other experiments, we set $num_{\text{sample}}=4$ due to our 4-process parallelization; $num_{\text{sample}}\geq 6$ yields similar accuracy with slightly more time.
    \item Effect of $num_{\text{refine}}$: This parameter upper-bounds the number of refinement iterations. We observe similar performance stability for $num_{\text{refine}} \geq 6$. In our other experiments, we set $num_{\text{refine}}=7$ to a relatively large number to make full use of the time budget for more refinement iterations; $num_{\text{refine}}=5$ yields higher efficiency with similar accuracy.
\end{itemize}
Overall, we do not observe significant sensitivity of \parf to these hyper-parameters (except for small values of $num_{\text{sample}}, num_{\text{refine}}$). The specific configuration of hyper-parameters (marked by dashed lines in \cref{fig:hyper-parameters}) in our other experiments is not finely tuned.
}

\subsection{\revision{RQ3: Generality of \parf}}
\label{sec:experiments-generality}

\revision{
In this experiment, we demonstrate the generality of {\parf} by interfacing it with another static analyzer {\mopsa}~\cite{DBLP:conf/vstte/JournaultMMO19}.}
\cref{tab:initial-probability-of-mopsa} collects the considered set of external parameters of {\mopsa} and their initial distributions used in {\parf}. Note that the parameter \texttt{-numeric} is a string-typed parameter with 3 possible values and thus is modeled as an integer-typed parameter (via a Poisson distribution) rather than the Boolean type. We use the same hyper-parameter values as for \eva except that we set $\textit{num}_{\text{refine}}=3$ to ensure that most analyses can terminate within the time budget (since \mopsa runs much slower than \eva on the OSCS benchmarks).


\revision{
The last two columns of \cref{tab:experiment-results} demonstrate how our \parf strategy improves the performance of \mopsa.\footnote{\revision{The \official and \expert strategies are unavailable in this experiment since \mopsa provides neither official parameter settings for OSCS benchmarks, nor built-in precision levels (like \texttt{-eva-precision} in \eva) for applying the \expert strategy.}} Observe that \mopsa fails to analyze some OSCS benchmarks which either run out of time (marked by TO) or have language features, data types, and/or symbols not supported by \mopsa (marked by --). For the rest benchmarks, \parf achieves the best results on 26/27 (96.3\%) program repositories with exclusively best results on 12/27 (44.4\%) cases, which significantly outperforms \mopsa's \default strategy. 
}

\revision{
Overall, this experiment demonstrates that \parf can be generalized to improve the performance of other static analyzers.
}

\subsection{RQ4: Improving \framac in SV-COMP}


\cref{tab:parf-in-SVCOMP} shows that {\parf} can slightly improve the performance of {\framac} in SV-COMP. Since the analysis resource for each verification task is limited to 15 minutes of CPU time, ${\framacsv}_{\text{precision11}}$ strategy uses a fixed highest \texttt{-eva-precision 11} parameter for analysis. We set the experimental strategy of ${\framacsv}_{\parf}$ as: (i) using \parf to find a high-accuracy parameter setting in the first 7.5 minutes, and (ii) pass this parameter setting into Frama-C for analysis in the second 7.5 minutes. The verification results and scoring conditions are classified as: (i) correct (i.e., the analysis result is consistent with the answer), get 1 or 2 points; (ii) wrong (i.e., the analysis result is opposite to the answer), deduct 16 or 32 points; (iii) unknown (i.e., the analysis result contains alarms that may be false positives), no change in score; (iv) failure (i.e., the analysis exceeds the time limit), no change in score.

\begin{table}[t]
    \caption{SV-COMP verification results in terms of \textbf{RQ4}.}
    \label{tab:parf-in-SVCOMP}
    \centering\small
    \begin{tabular}{lccccc}
        \toprule
        \multirow{2}{*}{\textbf{Setting}}  & \multicolumn{4}{c}{\textbf{Verification Result}} & \multirow{2}{*}{\textbf{Score}} \\
        \cmidrule(lr){2-5}
            & correct & wrong & unknown & failure &  \\
        \midrule
        ${\framacsv}_{\text{precision11}}$ & 146 & 3 & 272 & 33 & 186\\
        ${\framacsv}_{\parf}$ & \textbf{151}  & 3 & 300 & \textbf{0} & \textbf{196} \\
        \bottomrule
    \end{tabular}
\end{table}

The experimental results show that {\parf} can eliminate all analysis failures and successfully verify 5 more tasks, thus slightly improving the total score from 186 to 196. This is because (i) Most analyses using \texttt{-eva-precision 11} parameter of the verification tasks under the NoOverflows category can terminate within 15 minutes, as discussed in \cref{sec:parf-selecting-strategies}, so it is difficult for {\parf} strategy to obtain more accurate analysis results; (ii) {\parf} adaptively finds 5 high-accuracy analysis results among 33 ${\framacsv}_{\text{precision11}}$ analysis failures, thus successfully verifying these 5 tasks.

%% file: body/limitations.tex
\section{Limitations and Future Work}\label{sec:limitations}

We pinpoint several scenarios for which the proposed {\parf} framework is inadequate and provide potential solutions thereof.

First, {\parf} treats the underlying static analyzer as a black box. Although this feature facilitates the portability and generality of {\parf} (as discussed in \cref{sec:Methodology}), it does not exploit analyzer-specific functionalities and/or internal results, e.g., the data/control-flow analysis and weakest precondition reasoning of {\framac}~\cite{DBLP:journals/fac/KirchnerKPSY15} to further improve the quality of the generated parameters. It is therefore worth investigating to what extent we can enhance parameter refining by leveraging a \emph{white/grey-box model}, as is the use of \textsc{Sparrow}~\cite{10.1145/2254064.2254092} in \textsc{Bingo}~\cite{raghothaman2018user}. As an example, one may utilize or even manipulate the call graph of a program -- as an intermediate result of a static analyzer -- to achieve more fine-grained parameter tuning, e.g., setting analysis parameters for individual functions of a program (aka, the problem of \emph{function-level refining}).

Second, {\parf} does not employ the syntactic or semantic characteristics of the source program, e.g., the maximum number of loop iterations. However, we foresee that machine-learning models and techniques may be developed to \emph{learn a good parameter setting} based on such characteristics, which can then be used as the initial (distribution of) parameter setting of {\parf}.

Third, {\parf} models different parameters as independent random variables. Taking into account the dependencies between parameters is expected to reduce the search space and thereby accelerate the parameter refining process. To this end, we need to extend {\parf} to admit the \emph{representation of stochastic dependencies}.

%% file: body/relatedwork.tex
\section{Related Work}\label{sec:related-work}

\paragraph*{\bf Adaptive Parameter Tuning}
Automatic parameter-tuning techniques have witnessed numerous applications in many fields, including database~\cite{Database} and big data~\cite{ReviewTuning,7807262} systems, machine learning hyper-parameter tuning~\cite{ml1,ml2}, mathematical software~\cite{MindOpt}, and symbolic execution engines~\cite{chaSymTunerMaximizingPower2022a}. Our work is dedicated to providing automated parameter-tuning support for static analyzers.

We emphasize that our {\parf} framework is inspired by \symtuner~\cite{chaSymTunerMaximizingPower2022a}. \symtuner is an adaptive parameter tuning framework for symbolic execution tools based on a formulated discrete sample space of external parameters. It has an online learning algorithm to update probability distribution based on accumulated information. Whereas symbolic execution can stop at any time while still producing useful intermediate results, a static analyzer only yields complete alarm information until the whole analysis terminates.


Considering the inherent feature of abstract interpretation-based static analyzer, we have designed a novel representation of probability distributions and refinement strategy to ensure that the parameter probability distribution has both incrementality and adaptivity. These properties enable our approach to effectively avoid analysis failure and rewardless analysis. Therefore, \parf can complete more high-accuracy analyses within the time limit to obtain sufficient intermediate information for refinement.

\paragraph*{\bf Improving Static Analysis Tools}


Many static analysis tools have integrated strategies to address the parameterization problem, such as Astr{\'e}e~\cite{kastnerAbstractInterpretationIndustry2023} and \textsc{Goblint}~\cite{saanGoblintThreadModularAbstract2021}.
K\"astner et al.~\cite{kastnerAbstractInterpretationIndustry2023}
summarize the four most important abstraction mechanisms in Astr{\'e}e and recommend prioritizing the accuracy of related abstract domains, which amounts to narrowing down the parameter space. 
Kaestner et al.~\cite{kaestnerAutomaticSoundStatic2023} further present a mechanism to fine-tune the precision of Astr{\'e}e to the software under analysis to parameterize its abstract domains; however, this mechanism is not fully automated since it needs directives provided by the user.
Saan et al.~\cite{saanGoblintAutotuningThreadModular2023, saanGoblintAutotuningThreadModular2024} implement in \textsc{Goblint} a simple, heuristic autotuning method based on syntactical criteria, which can automatically activate or deactivate abstraction techniques before analysis. However, this method only generates an initial analysis configuration once and does not dynamically adapt to refine the parameter configuration.
Salvi et al.~\cite{salviTrueErrorFalse2014} present an approach to coupling model-based testing with static analysis based on a tool coupling between Astr\'ee and BTC\textsl{EmbeddedTester}\textsuperscript{\textregistered};
but this method cannot tune the parameters of the static analyzer. 

\paragraph*{\bf Probability-Based Algorithms}

In the past several years, researchers have presented numerous approaches to filter alarms of static analysis tools. These works leverage probabilistic methods to obtain prior knowledge of alarms, in order to determine whether a generated alarm is really caused by potential bugs instead of the upper approximation introduced by the analyzer.

Raghothaman et al.~\cite{raghothaman2018user} proposed \textsc{Bingo}, a system that utilizes Bayesian inference~\cite{Ackermann2019,InferenceOOPSLA24} to decide the confidence of each alarm. Their model is refined through ground truth labels provided by users in each iteration. Heo et al.~\cite{heo2019continuously} proposed \textsc{Drake}, a probabilistic framework to identify alarms relative to a certain program change, thus helping improve the efficacy of discovering true bugs. Chen et al.~\cite{chen2021boosting} presented \textsc{DynaBoost}. Their approach uses information obtained from test executions to bootstrap a system that ranks alarms and evolves with the help of user inspection, which extends the ability of the system beyond human-provided information. Kim et al.~\cite{kim2022learning} investigated the problem of being limited to certain models, which disables transferring knowledge from one program to another, and aggressively generalizing from external feedback, which leads to false negative results of classification. They proposed \textsc{BayeSmith}, which trains a Bayesian network to prioritize real bugs based on feedback effectively.

Different from our task, the above works mainly focus on checking the alarm list and determining true positives. They only run the static analyzer once and conduct no adjustments on input parameters, though both tasks aim to optimize the final analysis results.

%% file: body/conclusion.tex
\section{Conclusion}\label{sec:conclusion}

We have presented a novel framework called {\parf} for adaptively tuning external parameters of abstract interpretation-based static analyzers. {\parf} is -- to the best of our knowledge -- the first \emph{fully automated} approach that supports \emph{incremental} refinement of such parameters. The effectiveness of {\parf} has been demonstrated on a collection of standard benchmarks. Future directions include extending {\parf} to cope with function-level refining and dependencies between parameters, and encoding the algorithms into a probabilistic programming paradigm using Bayesian inference~\cite{Ackermann2019,InferenceOOPSLA24}.




